\documentclass{aa}  

\usepackage{graphicx}
\usepackage{txfonts}
\usepackage{xcolor}
\usepackage{lscape}
\usepackage{hyperref} 
\newcommand{\xmm}{{\em XMM-Newton}}
\newcommand{\chandra}{{\em Chandra}}
\newcommand{\nustar}{{\em NuSTAR}}
\newcommand{\rhoOph}{$\rho$~Oph}
\newcommand{\fxu}{{erg~s$^{-1}$~cm$^{-2}$}}
\newcommand{\lxu}{{erg~s$^{-1}$}}

\newcommand{\eltn}{Elias~29}

\begin{document}

  \title{A deep X-ray view of the Class I YSO Elias 29 with \xmm\ and \nustar.
  \thanks{Based on observations obtained with XMM-Newton, an ESA science mission with instruments and 
  contributions directly funded by ESA Member States and NASA}}

   \subtitle{}

   \author{I. Pillitteri
          \inst{1,2}
          \and
          S. Sciortino\inst{1}
          \and
          F. Reale\inst{1,3}
          \and 
          G. Micela\inst{1}
          \and 
          C. Argiroffi\inst{3}
          \and 
          E. Flaccomio\inst{1}
          \and 
          B. Stelzer\inst{4,1}
                    }

   \institute{INAF-Osservatorio Astronomico di Palermo, Piazza del Parlamento 1, 90134 Palermo, Italy\\
              \email{ignazio.pillitteri@inaf.it}
         \and
         Harvard-Smithsonian Center for Astrophysics, 60 Garden St., 02138 Cambridge MA, USA
         \and 
         Universit\`a degli Studi di Palermo, Piazza del Parlamento 1, 90134 Palermo, Italy
         \and 
         Eberhard Karls Universit\"at, Institut f\"ur Astronomie und Astrophysik, Sand 1, 72076 T\"ubingen, Germany
         }

   \date{Received ; accepted }

 
  \abstract
{ X-ray emission is a characteristic feature of young stellar objects (YSOs) and the result of  
 the interplay between rotation, magnetism and accretion. 
For this reason high energy phenomena are key elements to understand the process of star formation, 
the evolution of their circumstellar disks and eventually the formation of planets.
We investigated the X-ray characteristics of the Class I YSO Elias 29 with 
joint \xmm\ and \nustar\ observations of total duration 300 ks and 450 ks, respectively. 
These are the first observations of a very young ($<1$ Myr) stellar object in a band encompassing 
simultaneously both soft and hard X-rays ($0.3-10$ keV in \xmm, and $\approx 3-80$ keV in \nustar). 
The quiescent spectrum is well described by one thermal component at $\sim4.2$ keV absorbed by 
$N_H\sim5.5\times10^{22}$ cm$^{-2}$. In addition to the hot Fe complex at 6.7 keV, we observed 
fluorescent emission from Fe at $\sim6.4$ keV, confirming the previous 
findings. The line at 6.4 keV is detected during quiescent and flaring states and its flux is variable.
The equivalent width is found varying in the $\approx 0.15--0.5$ keV range. 
These values make unrealistic a simple model with a centrally illuminated disk and suggest a role of the 
cavity containing \eltn\ and possibly reverberation processes that could occur in it.
We observed two flares, with duration of 20 ks and 50 ks, respectively but only the first flare 
was observed  with both \xmm\ and \nustar. For this flare, we used its peak temperature and timing as 
diagnostics to infer a loop size of about 
$1-2$R$_\odot$ in length, which is about $20\%-30\%$ of the stellar radius. This implies a relatively compact structure. 
We systematically observed an increase of $N_H$ during the flares of a factor five. This behavior has been observed 
during flares previously detected in \eltn\ with \xmm\ and ASCA. 
The phenomenon hints that the flaring regions could be buried under the accretion streams and at high stellar 
latitudes, as the X-rays from flares pass through gas denser than the gas along the line of sight of the quiescent corona. 
In a different scenario, a contribution from scattered soft photons to the primary coronal emission
could mimic a shallower $N_H$ in the quiescent spectrum. 
In the spectrum of the full \nustar\ exposure, we detect hard X-ray emission in the band $\approx20-80$ keV 
in excess with respect to the thermal emission and significant at a level of $\geq2\sigma$. 
We speculate that the hard X-ray emission could be due to a population of energetic electrons accelerated 
by the magnetic field along the accretion streams.  These particles could concur to pumping up the Fe fluorescence 
when hitting cold Fe of the circumstellar disk along with X-ray photons with $E>7.11$ keV. 
}

   \maketitle
%

\section{Introduction}

X-ray observations of star-forming regions (SFRs) have established young stars
as bright X-ray sources, from the Class~I stage, when a thick envelope shrouds the
central object, through Class~II, when a thick disk has been fully formed and is visible, to the Class~III stage, 
where very little, if any, circumstellar disk or
envelope remains, the accretion process has ceased, proto-planets may have formed and the photosphere of the
disk-less star is hardly distinguishable from that of more mature stars
\citep{Montmerle1990,Feigelson1999,Favata2003}

Extensive and deep surveys of SFRs in X-rays have been obtained  with \chandra\ and \xmm\ (eg. COUP, XEST, DROXO, CCCP, \citealp{Getman2005,Guedel2007a,Pilli+2010,Townsley2011}). From these data we assessed that a large fraction of 
the X-ray emission of Class~I and II Young Stellar Objects (YSOs) 
is of coronal origin as clearly shown, for example, by impulsive activity similar to the flares 
observed in the solar corona. The magnetic structures that form the stellar coronae of YSOs sometimes can create
rotationally modulated emission  \citep{Flaccomio+05}. Another component of the X-ray emission 
likely arises from to the interaction of the central star and its circumstellar disk. This can be due
to infalling matter heated by the accretion process (e.g. \cite{Kastner2002}).   
A coronal activity affected by the accretion process has been proposed for explaining the soft X-ray excess observed 
in young accreting stars \citep{Guedel2007}. Another phenomenon is the fluorescent emission, mostly in the
neutral Fe  of the disk at 6.4 keV and likely stimulated by coronal X-rays with energies $>7.11$ keV 
\citep{Imanishi+2001,Tsujimoto+2005}
The YSO YLW16A in the $\rho$ Ophiuchi Dark Cloud is where for the first time during a large flare \cite{Imanishi+2001} detected
  prominent Fe K$\alpha$ line at 6.4 keV in its \chandra\ spectrum. 
They explained the feature as the result of the excitation of neutral Fe from hard X-ray photons produced during the flare.
In the spectra obtained with a 850 ksec long continuous ACIS observation dubbed the \chandra\ Orion Ultradeep Pointing 
(COUP) \cite{Tsujimoto+2005} reported the detection of a Fe  K$\alpha$ 6.4 keV line in seven flaring
sources in Orion. The 500 ks Deep \rhoOph\ \xmm\ Observation (DROXO) of the core F region
revealed 61 $\rho$ Ophiuchi YSO members \citep{Pilli+2010}. 
In nine of these 61 YSOs, specifically in 4 Class~I, 4 Class~II, 1 Class~III objects, \cite{Stelzer+2011}
detected the Fe K$\alpha$ 6.4 keV line both during flaring and quiescent phases. 

By using a novel Bayesian method, \cite{Czesla+Schmitt2010} reanalyzed the COUP data and  
found the Fe K$\alpha$ 6.4 keV line in 23 out of 106 YSOs in the Orion Nebula.
From these results we can infer that fluorescence occurs more frequently than previously thought. 
In some cases the emission is associated with soft X-ray flares, but it sometimes appears as a steadily
persistent feature, even during quiescent periods. 

\eltn\ is a Class~I/II YSO in the Rho Oph Dark Cloud where \cite{Giardino+2007} (hereafter Paper I) 
detected significant variability in the equivalent width (EW) of the Fe K$\alpha$ 6.4 keV line 
during the DROXO observation. 
The 6.4 keV line was weak, but present, during the first quiescent time interval 
(cf. Fig. 4 of Paper I, EW$\sim30$ eV) and  appeared at its maximum strength 90 ks after Elias 29 underwent a flare  with 
an EW$\sim800$ eV. The thermal X-ray emission was the same in the two time intervals, while variability of the 6.4 keV line 
was significant at a 99.9\% confidence level.

As for the excitation mechanisms, photo-ionization alone could not be sufficient to explain strong fluorescent emission
with EW in excess of $\sim150$ eV and other mechanisms like collisional excitation are invoqued. 
\citet{Drake2008} analyzed the Fe K$\alpha$ fluorescent line emission in a few stars concluding 
that there was not compelling evidence for a collisionally excited fluorescence from high energy electrons. 
On the one hand a simple disk illuminated geometry cannot produce EW in excess of 150 eV and 
thus the origin of the strong emission observed in Elias 29 is still not clear. More in general \citet{Drake2008} has 
considered four different possible explanations for the case of Fe K$\alpha$ with EW$>$150 eV, namely: 
1) high Fe abundance of the disk material that could increase line intensity, but this rapidly saturates at EW $\sim$ 800 eV 
\citep{Ballantyne+2002}, 
2) disk flaring that, thanks to a favorable geometry, can increase line intensity by a factor two or three; 
3)  emission induced by an "unseen" flare obscured by the stellar disk implying that the evaluation of the exciting continuum is 
grossly underestimated; 4) excitation due to high energy non-thermal electrons that however requires a substantial amount of 
energy stored in the impinging particles \citep{Ballanyne+Fabian2003}.
Since the presence of the Fe K$\alpha$ fluorescent line with EW$>$150 eV is a quite common feature among YSOs, 
explanations based on ad-hoc geometry of the system or peculiar conditions of the systems seem still unsatisfactory.
The extraordinary example of Fe fluorescence of V~1486 Ori \citep{Czesla+Schmitt2007},  where an EW of $\sim$ 1400 eV 
has been measured, can be explained only recurring to an excitation due to highly energetic particles.
Notice however that fluorescent Fe emission at $\sim6.4$ keV is observed also in active galactic nuclei (AGN), 
where, similarly to what happens in young stars, a central X-ray source illuminates the cold material located in the surrounding torus/disk system. There large EW ($0.2<EW<2$ keV) are often observed, especially for 
source with N$_H > 10^{23}$ cm$^{-2}$ (e.g. \citealp{Fukazawa2011}).
The soft ($0.3-10$ keV) and the hard ($>10$ keV) X-rays spectrum of a YSO with disk showing Fe fluorescence
can reveal the presence of a non-thermal population of electrons responsible for at least part of the fluorescence.
In this context we obtained a joint and simultaneous \xmm and \nustar\ observation of \eltn\ devoted to acquiring
spectra from soft (\xmm\ band 0.3-8.0 keV) to hard (\nustar\ band 3-80 keV) X-rays. 
We conceived this program in order to detect any non-thermal hard X-ray emission from \eltn, 
study the time variability and relate these features to the fluorescent emission, and eventually explain 
its origin.

We present the characteristics of the new X-ray observations and the adopted analysis in Sect. \ref{observations}, 
we illustrate the results of the time-resolved spectral analysis in Sect. \ref{results}, 
discuss the results in Sect. \ref{discussion}, finally we draw our conclusions in Sect. \ref{conclusions}.

\begin{figure*}[h]
\begin{center}
\resizebox{0.95\textwidth}{!}{
\includegraphics{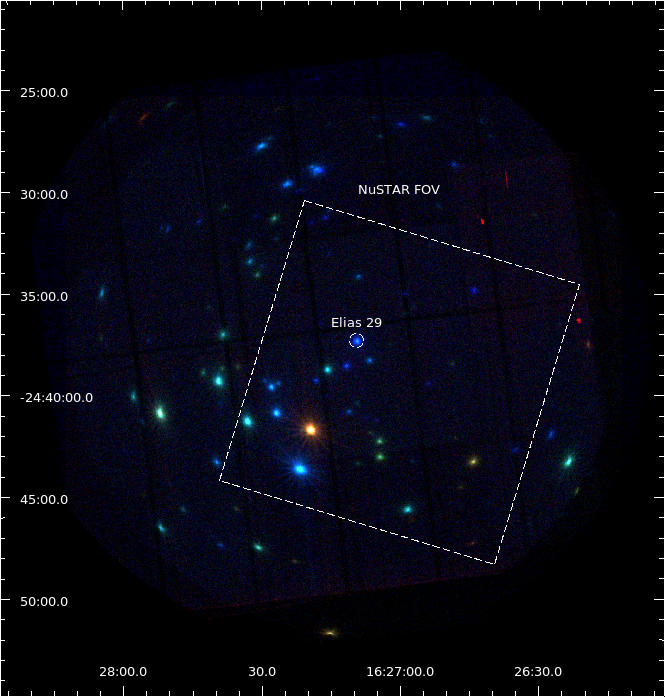}
\includegraphics{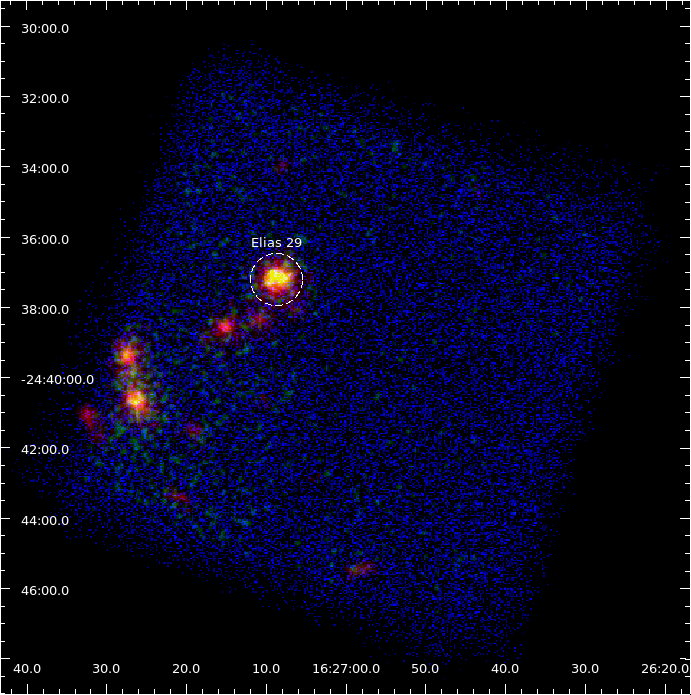}
}
\end{center}
\caption{\label{images} 
Left: color-coded image of the EPIC data integrated over the entire observation 
(red: $0.3-1.0$ keV; green: $1.0-3.0$ keV; blue: $3.0-8.0$ keV). The square indicates the field of view of the \nustar\ 
observation. 
Right: color-coded image of the \nustar\ data integrated over the entire observation 
(red: $3-10$ keV; green $10-40$ keV; blue $40-60$ keV). In both images a circle marks the position of \eltn.}
\end{figure*}

\section{Observations and data analysis \label{observations}}
\eltn\ ($\alpha=16^h27^m09.4s$, $\delta=-24^d37^m18.9^s$, other identifiers: [GY92] 214, 2MASS J16270943-2437187, 
ISO-Oph 108) is the most IR luminous Class I YSO in the Rho Ophiuchi Dark Cloud \citep{Bontemps2001,Natta2006}. 
Its accretion rate is about $1.5\times10^{-6}$ M$_\odot$ yr$^{-1}$, the circumstellar disk has an estimated mass of 
about 0.012 M$_\odot$, its inner radius is $\sim0.36$ AU which is about 13 stellar radii ($R_\star \sim 5.7-5.9 R_\odot$), and the outer radius of the disk is about 600 AU \citep{Boogert2002,Miotello2014,Rocha2015}. 
The system is tilted in a way that the line of sight partly crosses the envelope, 
the star is visible through the outflow cavity as well as a portion of the disk (cf. Fig. 14 in \citealp{Rocha2015}). 

The \xmm\ and \nustar\ observations were acquired as part of a large, joint program (PI: S. Sciortino). 
The total exposure time was $\sim 300$ ks for \xmm\ and $\sim 450$ ks for \nustar, but due to the low orbit of \nustar\ about 250 ks of science exposure were obtained with \nustar. 
The surveyed region covered most of the dense core F of LDN 1688, approximately 6\arcmin\ north of the 
previous pointing of the 500 ks long \xmm\ observation called DROXO \citep{Pilli+2010}.
Basic information on the \xmm\ and \nustar\ observations are reported in Table \ref{log}.  

\begin{table}
\caption{Log of the observations \label{log}. We will refer to the \xmm\ observations as first, second and 
third \xmm\ observation respectively. Analogously we will refer to the first, second and third \nustar\ 
observation for  simplicity.}
{\centering
\resizebox{\columnwidth}{!}{\begin{tabular}{lrrr}
\\\hline\hline
Satellite & ObsID  & Start (UT) & Net Exposure (ks)  \\
\hline
\xmm\   &   0800030801 (First) & 2017-08-13T16:34:58 & 99.65  \\
\xmm\   &   0800030901 (Second)& 2017-08-15T16:26:29 & 100.11 \\
\xmm\   &   0800031001 (Third)& 2017-08-17T19:26:33 & 95.02  \\\hline
\nustar\ &  30301001002 (First)& 2017-08-13T14:36:09 & 55.34  \\
\nustar\ &  30301001004 (Second)& 2017-08-15T14:56:09 & 94.24   \\
\nustar\ &  30301001006 (Third)& 2018-06-15T16:06:09 & 102.3   \\
\hline
\end{tabular}
}}\\

\footnotesize Note: for \nustar\ the science time per orbit is about 55\% of the orbit duration.
The third \nustar\ observation was obtained about 10 months after the joint \xmm and \nustar\ observations.
\end{table}

\subsection{\xmm\ observations}
The three \xmm\ observations have been carried out on three subsequent satellite orbits 
(orbits $3238$ to $3240$), with nomimal aim point at \eltn\ and very little variation of the position angle 
among the three orbits. We will refer to the \xmm\ observations as first, second and third \xmm\ observation respectively.
A log of the observations is reported in Table \ref{log}.
The \xmm\, EPIC ODF data were processed with SAS 
software\footnote{See \url{http://xmm.vilspa.esa.es/sas}} (version 16.1.0) and the latest calibration 
files in order to produce full field of view lists of events calibrated in both energy and astrometry
(Fig. \ref{images}).
 
We have subsequently filtered these photons and retained only those with energy
in the $0.3-8$ keV band and only the events that triggered simultaneously at most
two nearby pixels ({\sc FLAG==0; PATTERN <= 12}). 
This filtering was operated on the data of each EPIC detector (MOS1, MOS2, PN) 
and for each of the exposure segments of the three \xmm\ observations. 
The chosen energy band limits ensure a good overlap with the \nustar\ band 
and the best EPIC calibration. 

High background variability was present during the first part of each \xmm\ exposure. 
This has the effect to increase the noise in light curves and spectra when subtracting the background. 
However, for \eltn\ we preferred to use the full exposure time 
rather than excising the intervals with high background in order to have a continuous monitoring.
We used a circular region of radius $30\arcsec$ centered on the source centroid to extract the events 
for both MOS and PN. This region should contain about 80\% of the encircled energy of the \xmm\ 
point spread function (PSF).

The background events have been extracted from a nearby circular region of $40\arcsec$ radius without sources 
from the same chip and, for the PN, at the same distance from the read out node, as prescribed by the SAS guide. 
In order to produce the spectra we used a more strict selection (PATTERN <=4) as recommended in the SAS guide. 
With SAS we obtained light curves and spectra for both source and background events, response matrices (RMF)
and effective area files (ARF) for the spectral analysis. The  spectra were grouped to have at least 25 
counts per bin before the analysis with the {\sc XSPEC} software.

\subsection{\nustar\ observations}
Two \nustar\ observations were taken simultaneously with the \xmm\ observations, 
while a third exposure of duration $\sim195$ ks was obtained in June 2018. 
This third exposure without an \xmm\ counterpart was not initially planned as part of the campaign.  
Due to the low satellite orbit, the total of the science exposures amounts to $\sim250$ ks out of a total exposure 
of $\sim 450$ ks. 
The \nustar\ data were processed with the {\sc heasoft} suite (version 6.22.1), 
the \nustar\ dedicated pipeline and the latest calibration files (CALDB ver. 4.8) in order to 
produce full field of view lists of events calibrated in both energy and astrometry for the two cameras FPMA and FPMB. 
The resulting image in the $3-78$ keV band is shown in Fig. \ref{images} (right panel) where about 10 sources are recognized
by eye. 
The sources in the FOV are rather weak, with \eltn\ being the strongest one.

We adopted the standard thresholds for the rejection of particle background and the cut-off threshold at the 
SAA passage.
For \eltn\ we extracted the spectra from a circular region centered on the source centroid and of 
radius of 40$\arcsec$, this region should contain about 40\% of the total source counts \citep{Harrison2013}; a circular area 
of 80$\arcsec$ radius was used to extract the background events from a nearby region.
The tasks {\sc nuproducts} and {\sc nupipeline} were used to extract events in different energy bands and
time intervals, and to create spectra, light curves and related calibration files like response matrices 
and {\sc arf} for the spectral analysis with {\sc xspec}.

\subsection{Spectral analysis}
Spectra from the  events of \xmm\ MOS1, MOS2 and PN and \nustar\ FPM A and B were accumulated in 
different time intervals for time resolved spectroscopy (see Sect.  \ref{timresolvspec}). 
The spectra (energy band 0.3--8.0 keV) were modeled with an absorbed thermal APEC component in
order to derive the properties of the emitting plasma, specifically the temperature, the emission measure (EM), 
the global abundances ($Z/Z_\odot$) and the flux. In addition, we used a Gaussian line with intrinsic width equal to zero
to model the fluorescent emission in the $6.4-6.6$ keV range \footnote{In XSPEC terminology {\sc TBabs(APEC+Gaussian)}}. 
The EPIC spectral resolution is the main factor of the broadening of the Gaussian line width.
In principle we could expect a variation of the lines contributing to the blend of fluorescent emission and their
relative strengths, however we do not expect velocity fields that can increase the line width to a detectable level.
The global abundance was derived from the best fit modeling to the spectrum
of the third \xmm\ observation (ObsId 080031001) with a value $Z/Z_\odot = 0.54$.  
This is consistent with sub-solar $Z/Z_\odot$ often derived 
from the analysis of low resolution spectra of young coronae \citep{Maggio+2000,Guedel2003,Maggio2007}.
For the other time intervals we used a fixed $Z/Z_\odot = 0.54$.

The results of this analysis are presented in Table \ref{bestfit}.
Overall, the spectrum of \eltn\ is smooth and the only prominent feature is at $6.4-6.7$ keV due to the \ion{Fe}{XXV} line 
and the neutral Fe line at 6.4 keV. 
The plasma temperature is found in excess of 50 MK ($\sim4.2$ keV), this means that most of the metals are completely
ionized and the only discrete features in the thermal spectrum are due to the hot \ion{Fe}{xxv} around 6.7 keV.
The blend between the 6.7 keV Fe line and the fluorescent line at $\sim6.4$ keV requires a careful estimate
of the APEC abundance and the temperature for avoiding biases in the line centroid and strength of the 6.4 keV 
line.
The determination of the centroid position and the Gaussian EW are affected by the low spectral resolution of 
EPIC, the limited count statistics of the spectra, the strength of the line, but also by the gas absorption, the
temperature and Fe abundance for the estimate of the continuum and the blend with the 6.7 keV line.
For this reason we also performed a best fit modeling to the  spectra in the energy range $5-8$ keV. 
In this narrower band the value of $N_H$ absorption is less constrained, however 
the continuum of the line is determined by the APEC temperature and its 
normalization, and the abundance value $Z/Z_\odot$  constrains the intensity of the 6.7 keV Fe line. 
The best fit value of the temperature found in the full band produce an acceptable fit in the narrow band, however only
for the purpose of the best evaluation of the continuum around 6.4 keV we let free to vary the temperature free and the 
global abundances. The results from such fits are listed in Table \ref{fit5-8} and we will refer to these results
for discussing the fluorescence (Sect. \ref{fluorescence}).  
The $N_H$ value was kept fixed to the value estimated from the quiescent level ($N_H=5.5\times10^{22}$ cm$^{-2}$).
Approximately, half of the counts in the full band spectrum  are present in the 5-8 keV band. 
For the first and third \xmm\ observations (ObsId 08030801 and 08031001) we considered the full exposure 
in order to keep an adequate level of count statistics, while for the second observation (ObsId 08030901) 
we analyzed the spectra in the same time intervals used for the
full energy band. The choice of the time intervals is detailed in Sect. \ref{timresolvspec}. 
 
\section{Results}
\label{results}
Visual inspection of the \xmm\ images shows the presence of more than 100 sources, 
while in the smaller \nustar\ FOV we can easily recognize about ten sources. 
We defer the study of the remaining X-ray sources
to a separate paper, while in this paper we  focus
on \eltn\, which is the strongest of the \nustar\ sources (cf. Fig. \ref{images}).

\subsection{\xmm\ and \nustar\ light curves of \eltn }

\begin{figure*}
\centering
\resizebox{\textwidth}{!}{\includegraphics{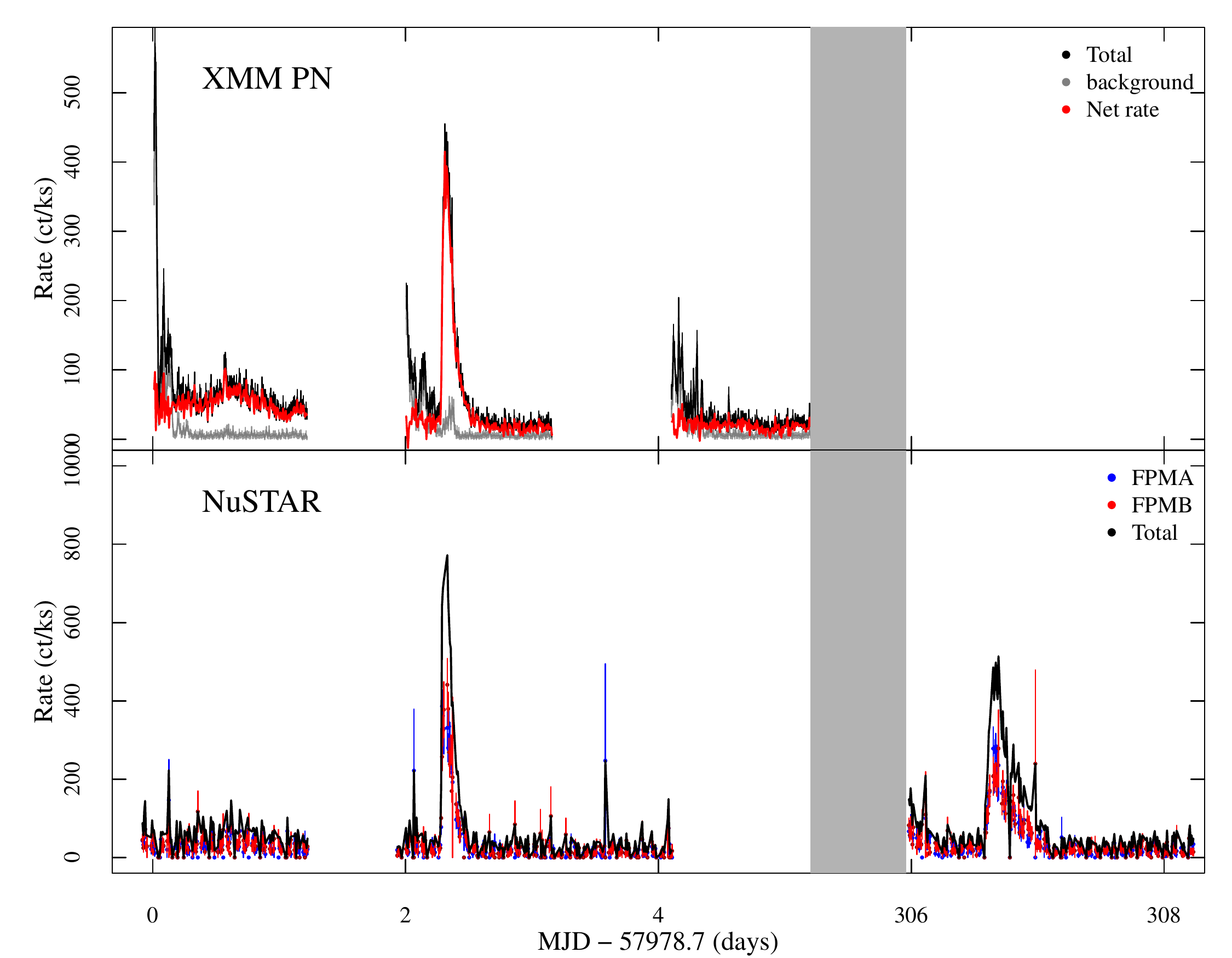}}
\caption{\xmm\ PN (top panel) and \nustar\ (bottom panel) light curves of \eltn. 
High background variability affects the first part of each \xmm\ exposure and during the main flare, 
but its effect can be adequately corrected as it is shown by the background-subtracted (red) light-curve. 
The origin of time axes is set to the time of start of the first PN exposure. The time gap between the second
and third \nustar\ exposures is marked with the gray area and amounts to about 300 days.}
\label{lcs}
\end{figure*}

The PN and FPM light curves or \eltn\  are shown in Fig. \ref{lcs}. 
Two major flares occurred during the exposures, but only the first flare was observed  
simultaneously with \xmm\ and \nustar.
The first flare had a duration of about 20 ks and a esponential decay time of about 7.6 ks, 
the second flare had duration was about 50 ks with an exponential decay time of about 9.3 ks. 
Before the second flare possibly the final decay of another flare was recorded with \nustar.
The lack of \xmm\ simultaneous coverage limits the information we can obtain about the second flare, 
however its detection allows to infer that 
approximately every 200-250 ks a flare of intensity similar to those observed in the present data 
can occur on \eltn. 
A detailed analysis of the first flare is given in Sect. \ref{flare}.
Other than the two main flares, we notice that in the first \xmm\ observation
(ObsID 080003081) the rate smoothly increased and then decreased on a time scale of about 50 ks, it also
showed a very short spike in the same time interval. 

\begin{figure*}
\begin{center}
 \resizebox{0.95\textwidth}{!}{
  \includegraphics[width=\columnwidth]{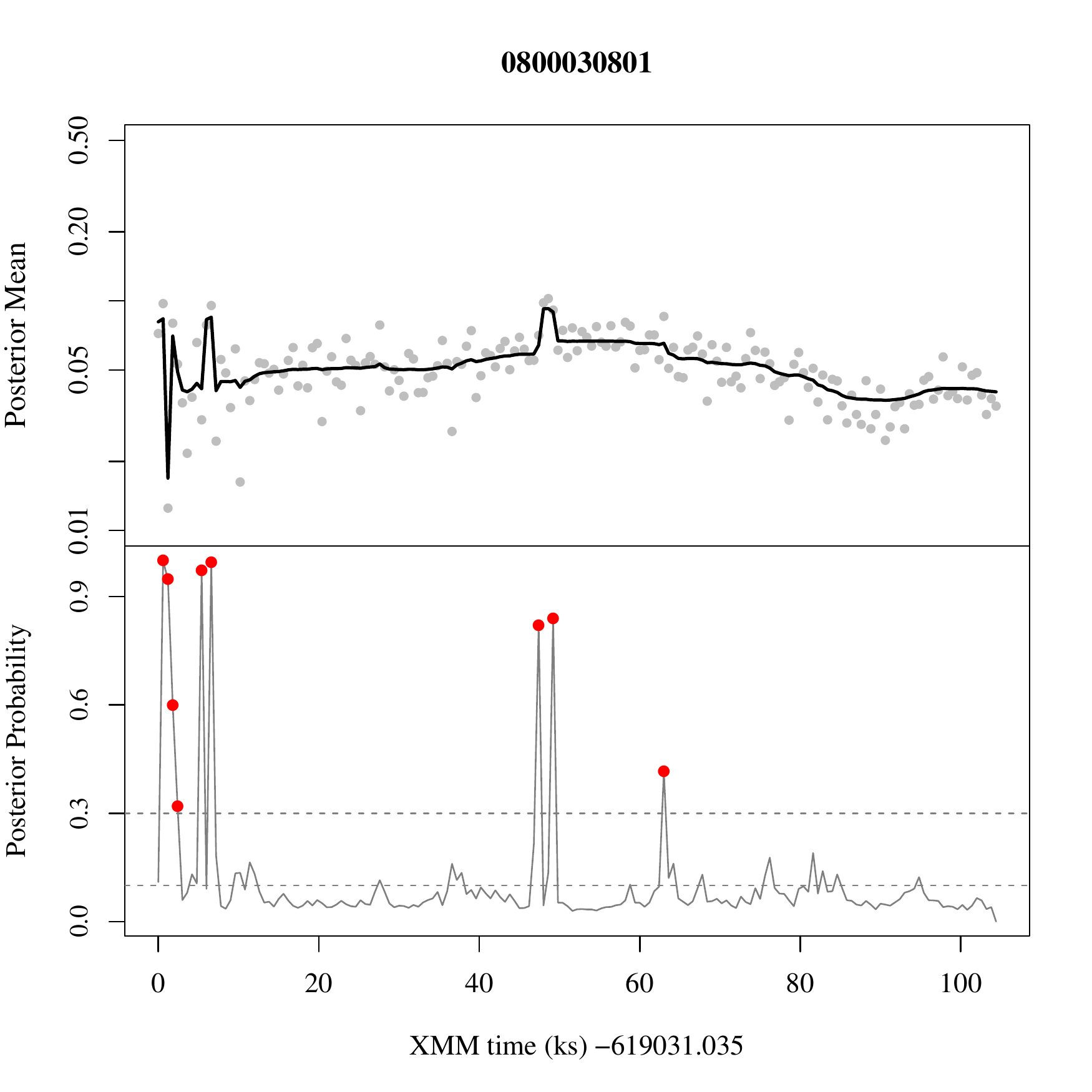}
  \includegraphics[width=\columnwidth]{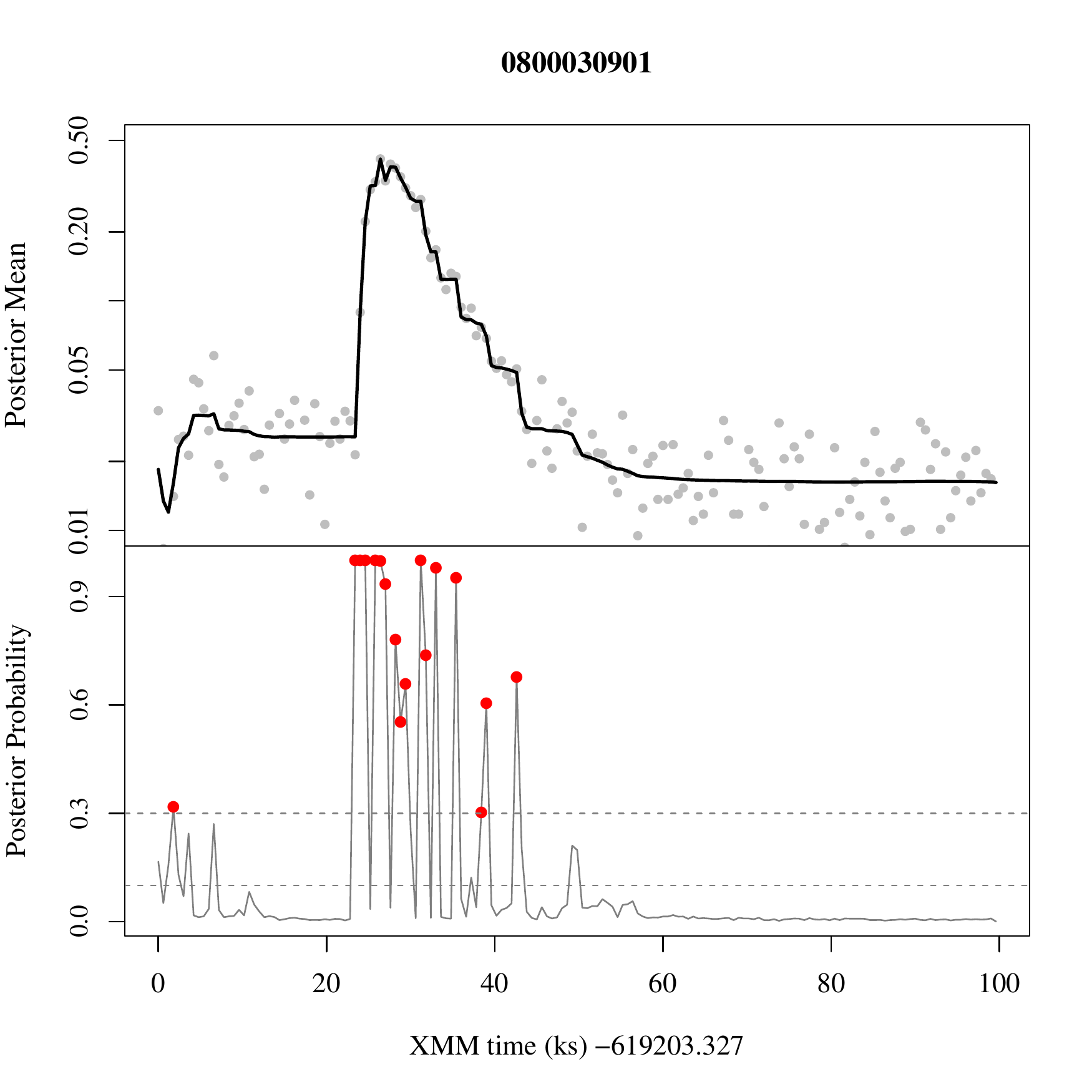}
  \includegraphics[width=\columnwidth]{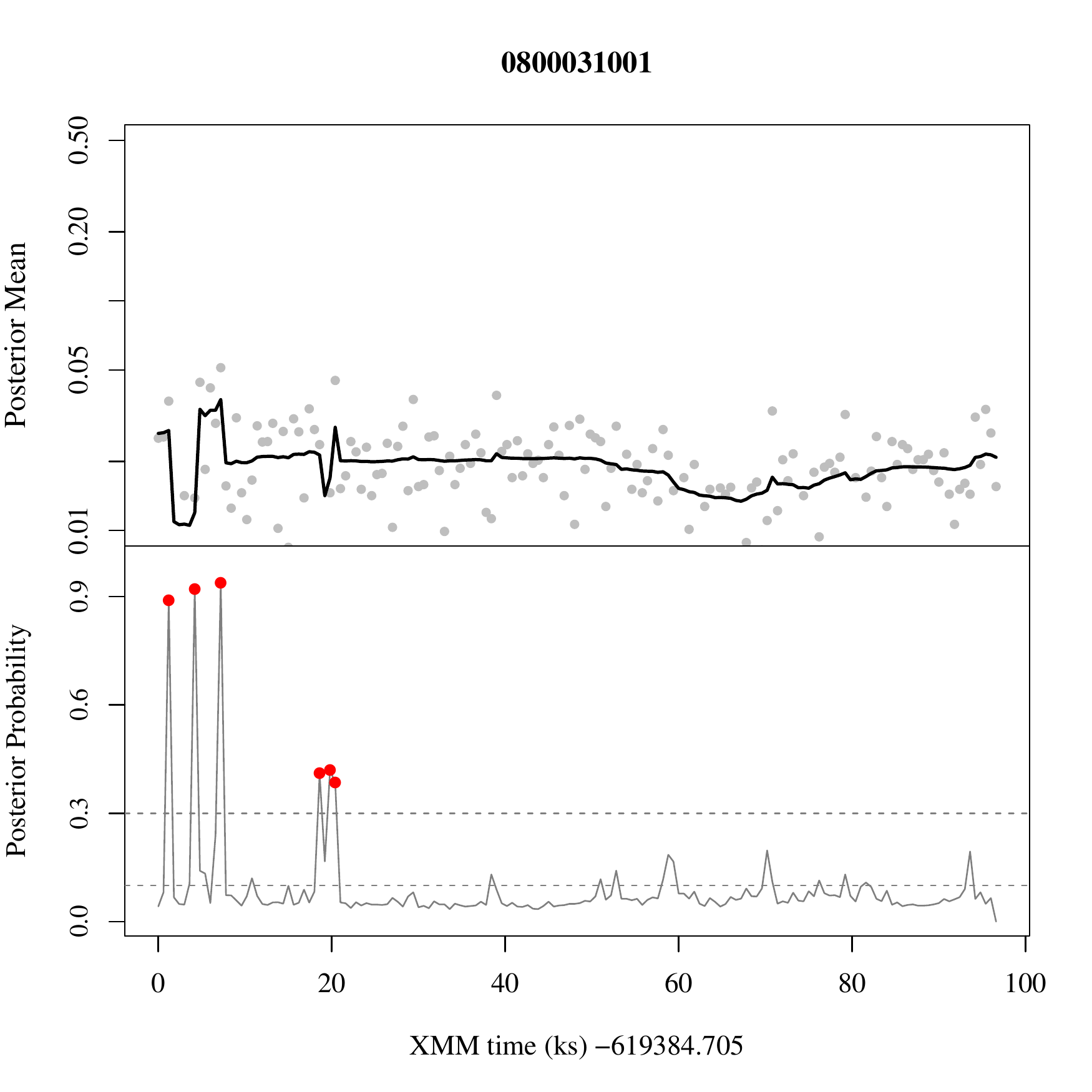}
 }
\end{center}
\caption{BCP analysis of the \xmm\ PN light curves in the 0.3-8.0 keV bandpass. 
Top panel shows the light curve (gray dots) with the posterior mean (solid line). The scale of the Y-axis is
logarithmic and with the same range of values across the 3 panels for ease of comparison.
Bottom panel shows the posterior probability at each point. We indicated
the probabilities $P>0.3$ with red dots, the horizontal dotted lines show the levels at
significance $P=0.3$ and $P=0.1$, respectively.}
\label{lc_bcp}
\end{figure*}

In order to study the time variability and identify any statistical change in the count rate, 
we compared two distinct techniques: the {\sl Bayesian change point (bcp)} analysis\footnote{Implemented in the $R$ package "bcp"} 
\citep{Wang+Emerson2015, Erdman+Emerson2007, Erdman+Emerson2008} and the {\sl Prune Exact Linear Time (PELT)} 
analysis\footnote{Implemented in the $R$ package "changepoint"} \citep{Killick+2012}.
 The first method (bcp) uses a bayesian approach to determine the change points in a time series. For each data point
it derives a posterior mean of the rate and a probability of change of the rate at each data point. 
The second method, PELT, uses a competitive algorithm that minimizes a cost function while guarding against overfitting
the data by means of a penalty function.
Fig.~\ref{lc_bcp} shows the posterior mean and the posterior probability of a change at each light curve bin obtained 
from the $bcp$ analysis.  
In the probability panel we can decide which threshold to use to identify variability at some level of significance.
For example, we can pick the values at $P=0.3$ and $P=0.1$, respectively. 
Above $P=0.3$ we can decide that there is a change of the rate, while between $P=0.1$ and $P=0.3$ we could have a likely change of the rate. 
Similar results are obtained for \nustar\ light curves, however the lower count statistics 
of the data introduces more spurious peaks of posterior probability above 0.3 and 
that do not look related to real variability. 

\begin{figure}
\begin{center}
\resizebox{0.5\textwidth}{!}{
\includegraphics[width=\columnwidth]{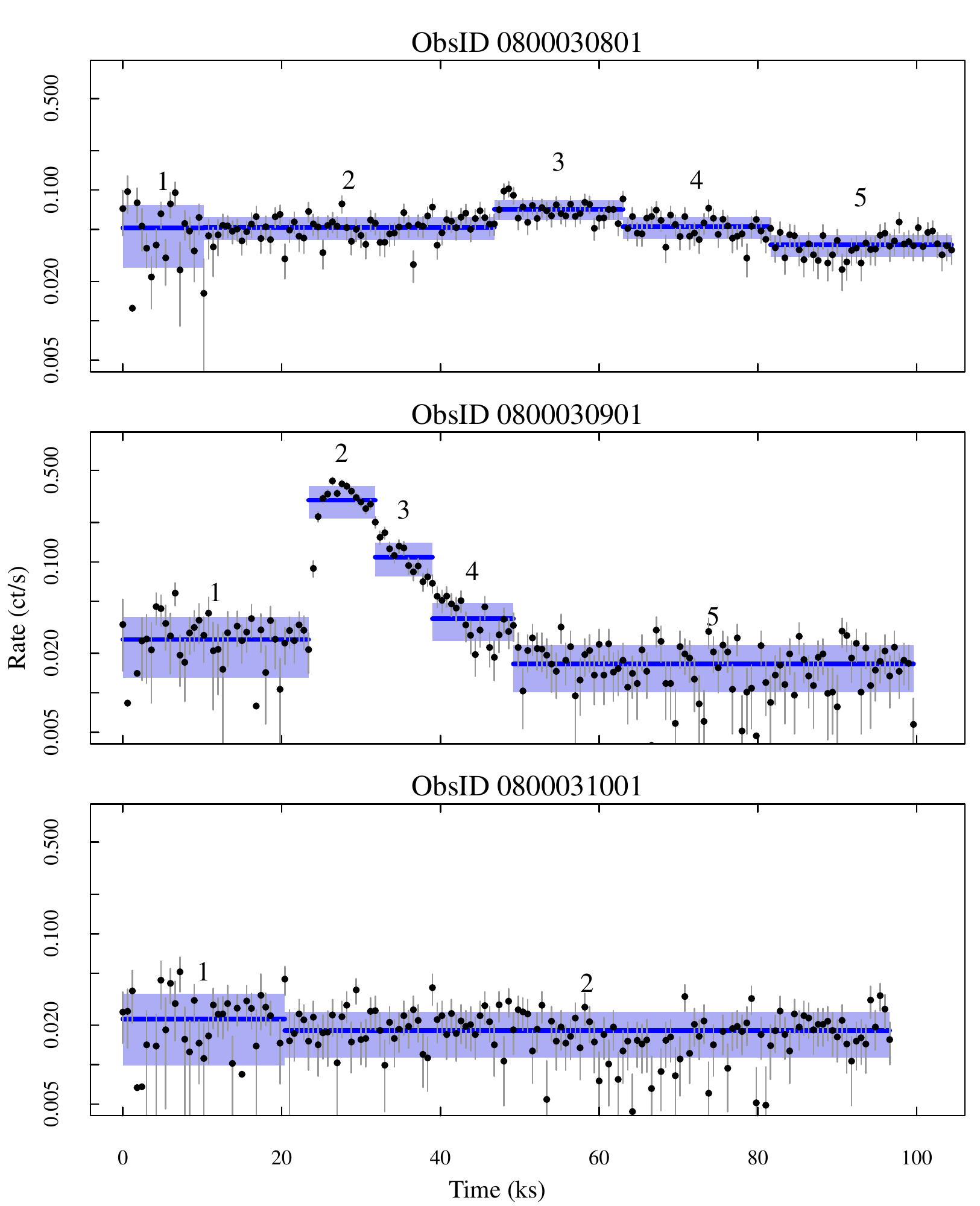}
}
\end{center}
\caption{\label{cpt_pn}
PELT analysis of the \xmm\ PN light curves in the 0.3-8.0 keV bandpass subtracted of background rate. 
Bin size is 600 s and the rate is displayed in a log scale. Panels have the same range on the y-axis. 
Horizontal segments and shaded areas mark the average count rate and variance in each time interval
(numbers on top of the intervals). The semi-log scale evidences the regular exponential decay of the flare.
}

\end{figure}

\begin{figure*}
\begin{center}
\resizebox{0.99\textwidth}{!}{
\includegraphics{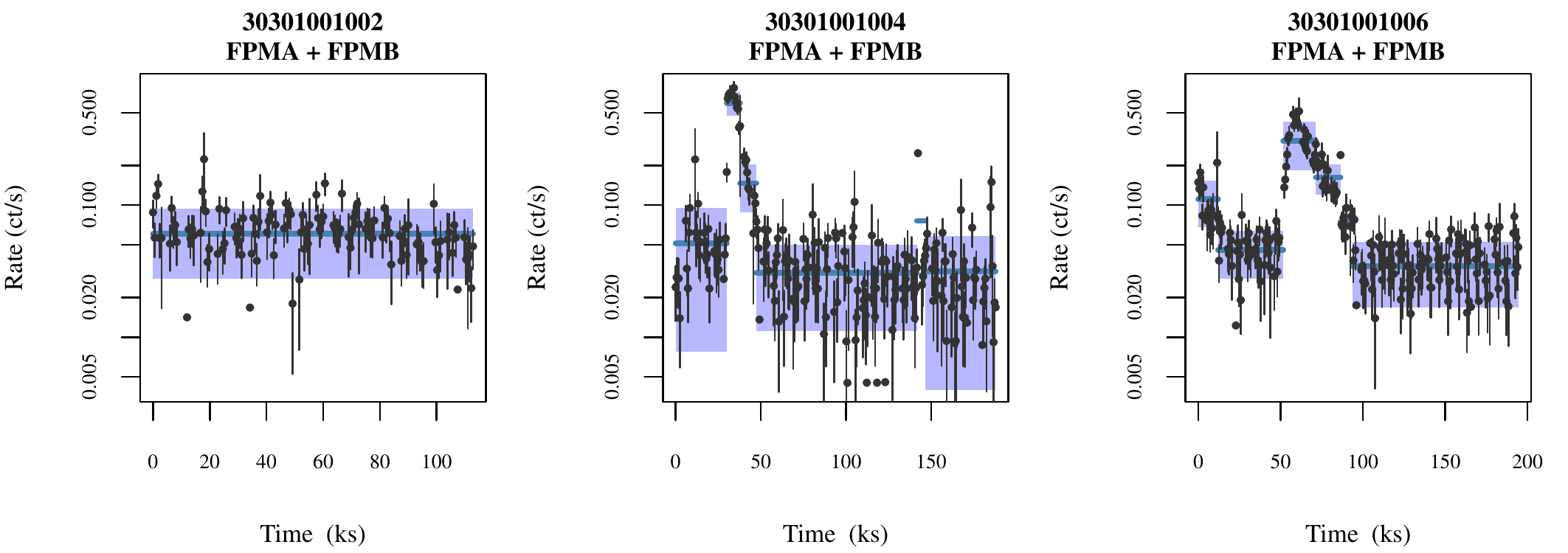}
}
\end{center}
\caption{ As in Fig. \ref{cpt_pn} for \nustar\ light curves in the $3-80$ keV. 
}
\label{cpt_nustar} 
\end{figure*}

Figs. \ref{cpt_pn} and \ref{cpt_nustar} show the results from PELT analysis: the input background-subtracted PN and 
FPM light curves are shown with over-plotted the time segments and individual segment average rate. 
We identified change points based on changes of mean rate and its variance ({\em cpt.meanvar} function).
We used an asymptotic type penalty and the default value of 0.05 (corresponding to a 95\% statistical 
significance level at each change point). 
We further checked the results of the number of intervals identified 
by using a {\em manual} value for the penalty function and producing a plot
of the number of change points as a function of the penalty\footnote{See, e.g.,   
\url{www.stats.stackexchange.com/questions/60245/penalty-value-in-changepoint-analysis/60440\#60440} }. 
Small values of the penalty produce more spurious change points, their number flattens out rapidly with increasing values of the penalty. The {\em elbow} corresponds to the number of expected change points.
Compared to {\sl bcp}  PELT seems less sensible to small variations of the rate
while {\sl bcp} analysis seems more capable to find short duration change of rate of smaller amplitude. 
On the other hand, the time segments found with PELT are more adequate for performing a robust time resolved 
spectral analysis as they include more counts overall.
We performed the same analysis on the narrow energy band of $6.1-6.9$ keV around the complex of the Fe lines
at $6.4-6.7$ keV. We obtained four intervals that identify the pre and post flare quiescent level, a flare peak 
segment and  a flare decay segment in agreement with the full energy band analysis.  
For the time resolved spectroscopy we used the time segments identified with PELT on the light curve of the full energy band
of \xmm\ (0.3-8.0 keV). 
The total net counts in the different intervals vary between $\sim600$ and $\sim3400$ (cf. Table \ref{bestfit}). 

The light curve of the third \nustar\ exposure is divided into five segments by PELT. We can recognize 
an initial partial decay phase, likely from an unseen flare, then a quiescent segment before the flare, a 
peak and a decay segment for the flare and the post-flare quiescent phase.

\subsection{\xmm\ time resolved spectroscopy \label{timresolvspec} }
\subsubsection{Quiescent emission} \label{quiet}

\begin{figure}
 \begin{center}
 \resizebox{1.00\columnwidth}{!}{
  \includegraphics[width=\columnwidth]{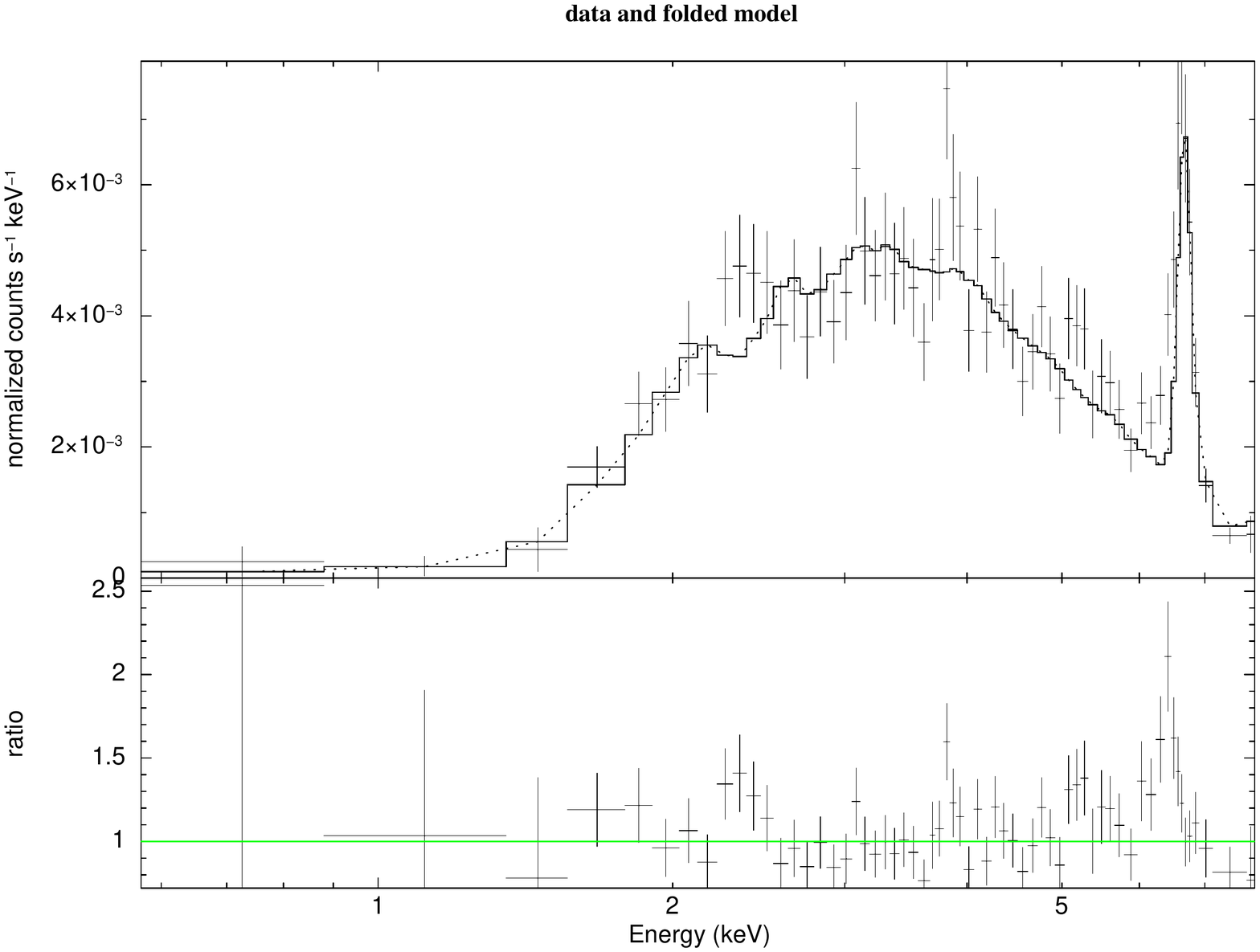}
 }
 \end{center}
  \caption{\label{qu_apec}  PN spectrum of the quiescent phase (top panel) and ratio data/model. The
  spectrum was accumulated from events collected during the quiescent phases after the flare and in 
  the third \xmm\ observation. The model  is an absorbed APEC thermal component. A sharp excess of emission 
  is visible at $\sim6.4$ keV and due to the fluorescent emission present even during quiescence.}
\end{figure}

The third \xmm\ observation shows a low PN rate for about 100 ks, and it appears as a  continued quiescent
phase after the flare registered in the second \xmm\ exposure. 
Despite PELT identifies two time intervals with different rate variance during the third exposure, we considered
the full exposure as a whole for producing the MOS and PN spectra. 
The PN spectrum is shown in Fig. \ref{qu_apec} with the best fit model composed by an APEC plus Gaussian line. 
The best fit parameters of the model are shown in Table \ref{bestfit}.
The average plasma temperature is $\sim4$ keV (90\% confidence range $3.1-5.6$ keV) and
the hydrogen column density $N_\mathrm{H}$  is $5.5\times10^{22}$ cm$^{-2}$ (90\% confidence range 
$3.1-5.6\times10^{22}$ cm$^{-2}$). These values are similar to those found by \citet{Giardino+2007} and by \citet{Favata+2005}, and thus we conclude that the X-ray coronal emission of \eltn\ has 
been stably hot over a time scale of $\sim12$ years. 
The quiescent unabsorbed flux in $0.3-8.0$ keV is about $6.3\times10^{-13}$ \fxu\ which corresponds to 
$L_X\sim 1.1\times10^{30}$ \lxu at 120 pc. 

Furthermore,  we accumulated a PN spectrum of the exposure encompassing the quiescent period after the flare 
in the second \xmm\ exposure (segment 5) and the whole third exposure.
This is justified by the similar count rate in both segments that suggests similar spectral characteristics 
of the plasma. 
The resulting PN spectrum had about 3400 counts, the best fit with an absorbed APEC component had values (90\% 
confidence range in braces): $N_H = 5.8 (5.3-6.4)\times10^{22}$ cm$^{-2}$, $kT= 4.2 (3.6-5.3)$ keV, $Z/Z_\odot = 0.6\ 
(0.4-0.8)$, $\log EM = 52.78 (52.77-52.79)$ cm$^{-3}$ and unabsorbed flux in $0.3-8$ keV band  $\log F_{APEC} = -12.21$ 
(-12.23  -- -12.19) \fxu (see Table \ref{bestfit}).
Fluorescence is firmly detected in the quiescent phase, as the modeling of the quiescent spectrum with a thermal 
APEC component alone shows a sharp excess of emission around 6.4--6.5 keV (Fig. \ref{qu_apec}). 
The Gaussian line gives a best fit centroid of 6.49 ($6.40-6.60$) keV, EW$=$0.25 ($0.16-0.38$) keV and 
flux of the line of $\log F_{Gau} = -14.17$ (-14.27 -- -13.93) \fxu. 


\subsubsection{Smooth variability}
\begin{figure}
 \begin{center}
 \resizebox{1.00\columnwidth}{!}{
 \includegraphics[width=\columnwidth]{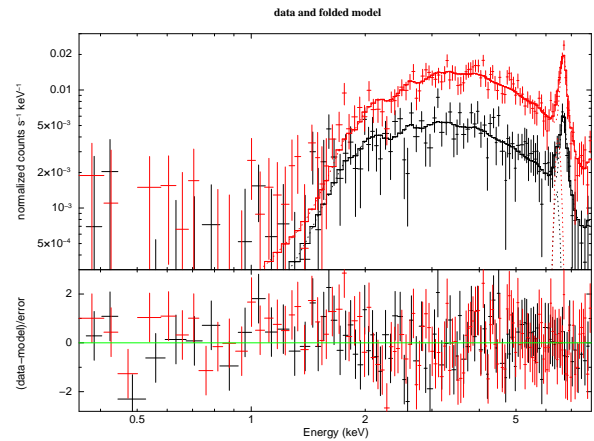}
 }
 \end{center}
  \caption{\label{fit_pn_1001}  PN spectra and best fit models { for the first \xmm\ exposure (red symbols) 
 and the third one (black symbols).}  Lower panel shows the residuals (data - model values). 
 The model for both spectra is an absorbed thermal APEC component plus a Gaussian to take into account the 
 fluorescent  emission from partially ionized Fe lines at $6.4-6.5$ keV. The difference between the two spectra
 is due to a difference of EM.}
\end{figure}

In the  first \xmm\ observation (ObsId 0800030801) the PN rate of \eltn\ showed a slow increase of the
rate followed by a similar smooth decrease. A spike of duration $\le2$ ks is visible 
near the center of the exposure,  beside this spike there is no evidence of other rapid variability. 
Following the subdivision in time intervals from PELT we performed time resolved spectroscopy in each
of the resulting five intervals with models as described in Sect. 2. 
The best fit values of the $N_H$ gas absorption vary in the range $5.4-7.2 \times10^{22}$ cm$^{-2}$ within 
the five segments, however these values are still consistent with each other and with the $N_H$ derived from the 
quiescent spectrum at a 90\% confidence level.
The best fit temperature, $kT$, varies in the $3.4-5.4$ keV range during the first observation, 
however kT values is still consistent with the quiescent temperature derived from the 
last \xmm\ observation at a confidence level of 90\%.
Also, the segment that contains the short spike has a somewhat high temperature (5.4 keV, 
90\% range $\sim4.2-6.9$ keV). 

In Fig. \ref{fit_pn_1001} we show the average spectrum of the { whole first exposure} of 
\xmm\ observation compared to the average spectrum of the { last exposure} which represents 
the quiescent emission.
While the gas absorption $N_H$ and plasma temperature kT were found similar in the two spectra 
($N_H\sim 5.5-6.5\times10^{22}$ cm$^{-2}$, kT$\sim 4-4.2$ keV) the EM was found larger by a factor of three 
during the first observation with respect to the EM of the third (quiescent) observation. 

 The smooth variability occurred on a time scale of less than one day and was not 
observed again in the following exposures. We speculate that the increase of rate  
could have been caused by an active coronal region appearing on view because of the
stellar rotation. Such a region would contain plasma denser than the rest of the corona thus resulting in 
an increase of the EM. In addition, the region could show flaring activity like 
the short spike we observed, as a result of the complex dynamics of the magnetic structures in it.
The passage of such a region lasted about 80 ks or $\sim0.93$ days, consistent with a rotation 
period of about two days, typical of a very young Class I YSO like \eltn. The fact that it did not appear
again in the second \xmm\ exposure sets a lifetime to the order of one day.
However, the region itself could have hosted the main flare observed in the second exposure when the region did 
appear again in view. The flare could have also destroyed or heavily reshaped the active region as this latter did not 
appear again in the third \xmm\ observation. 

\subsubsection{Flare analysis}\label{flare}
\begin{figure}
\begin{center}
\resizebox{0.95\columnwidth}{!}{
  \includegraphics[width=\columnwidth]{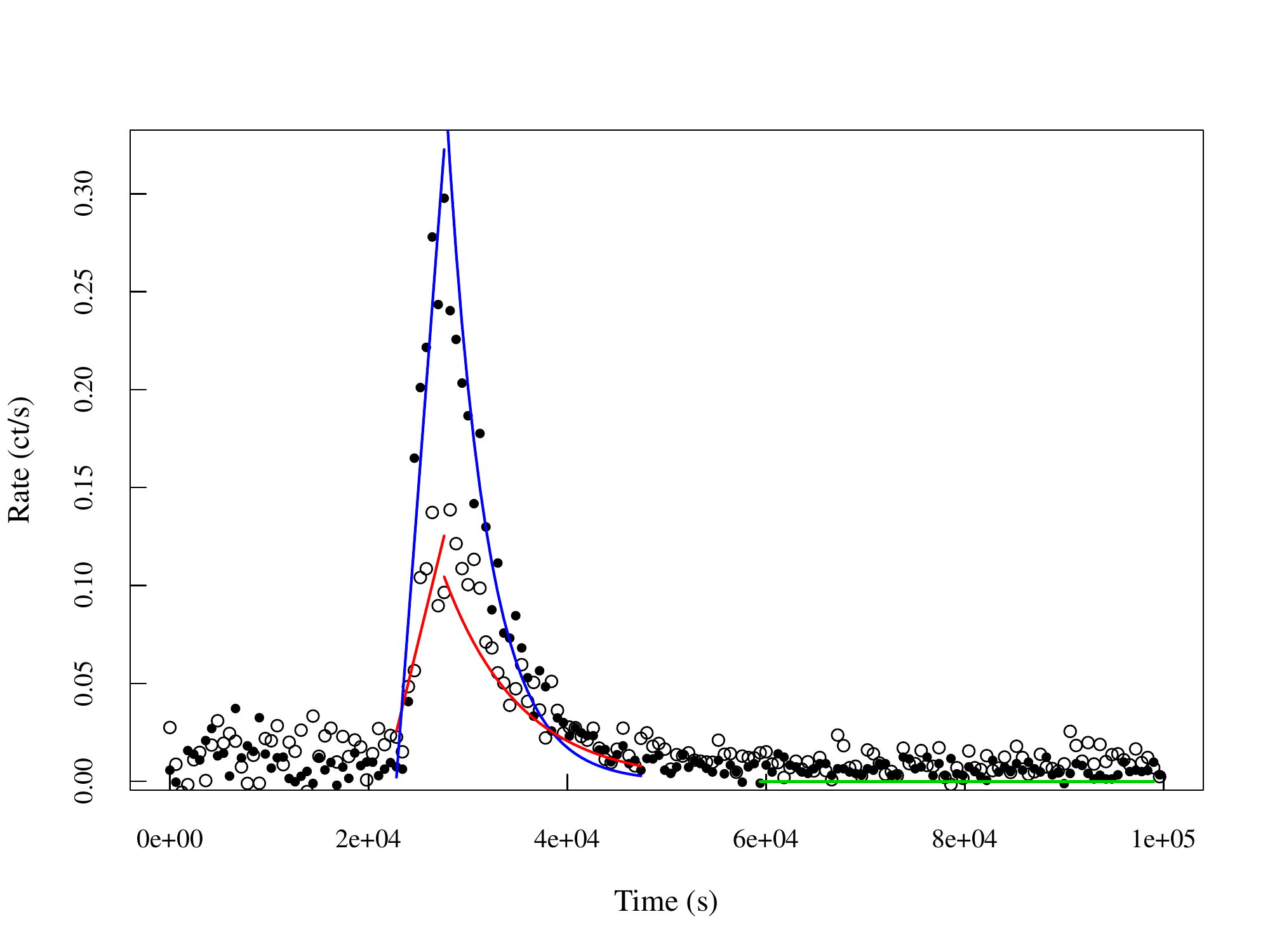}
}
\end{center}
\caption{\label{flare_timing}  \xmm\ PN light curves in the $0.3-5.0$ keV (open circles) and $5-8$ keV (solid dots) 
with the best fit of the rise phase (linear increase) and the decay phase (exponential decrease).
The rates are subtracted of the respective background rates.
The green segment marks the interval where the median of the quiescent rate has been
calculated. The quiescent rate has been subtracted to derive the decay times.}
\end{figure}

 Two main flares were observed in \eltn\ but we discuss in depth the first flare  because of  the simultaneous 
coverage with \xmm\ and \nustar\ coverage.
This flare showed a quite regular decay phase well modeled with an exponential decay. 
The decay of the flare appears faster in the hard band (5--8 keV) 
than in the soft band (0.3--5.0 keV), with e-folding times $\tau_0\sim 4.1$ ks in the hard band and 
$\tau_0\sim7.6$ ks in the soft band (Fig. \ref{flare_timing}).   
A similar timing is observed in \nustar\ data with a decay time equal to $\sim4.2\pm0.5$ ks and a rise time of
$\sim1.8\pm0.6$ ks.
{The light curve of the PN shows a peak rate $\approx8$ times the quiescent rate before the flare rise, with a peak luminosity $\log L_X = 31.18$. The duration and luminosity of this flare are larger than those of solar flares (e.g., \citealp{Tsuboi2016}) but not exceptional when compared to some of the flares observed in YSOs of the Orion Nebula 
\citep{Favata+2005, Caramazza2007, Wolk2005} and in $\rho$ Ophiuchi itself \citep{Pilli+2010}. }
A flare with similar duration and peak rate was observed by  \citet{Giardino+2007}. 
Taking into account the past X-ray observations, in a total exposure time of $\sim800$ ks \eltn\ has 
shown flares with duration of less than one day and peak rate about 10 times the quiescent rate. 
{ The second flare, observed only with \nustar, had total duration of about 50 ks (almost 14 hours), quite shorter than the day long lasting flares seen in the Orion Nebula and in $\rho$ Ophiuchi. }

The modest brightness of the flare implies a modest count statistics. This fact reduces the 
detail and accuracy of the time resolved spectroscopy we can perform on it.
The PELT algorithm divided the flare roughly in a peak segment (number 2), 
two decay segments (marked 3 and 4) and two quiescent segments (1 and 5) 
before and after the flare, respectively. Table \ref{bestfit} lists the 
best fit parameters of the flare segments. 
In order to improve the statistics and better constrain
the model parameters, we made a simultaneous fit of the spectra of segments 1 and 5 as they 
are representative of the quiescent phases before and after the flare.
In these time intervals we measured $N_H\sim7\times10^{22}$ cm$^{-2}$, $kT\sim5.9$ lkeV, $\log$EM$\sim52.8$ 
and unabsorbed  flux ($\log Flux \sim-12.19$), which are similar to the values obtained from the spectrum of the quiescent phase
of the third \xmm exposure.

During the flare there is an increase of both the temperature and the gas absorption. 
The temperature rises to about 11.1 keV ($\sim130$ MK) and the absorption reaches values
of $N_H \sim 2.1\times10^{23}$ cm$^{-2}$ which is about a factor of four higher than the $N_H$ of the
quiescent phase. The difference of $N_H$ between quiescence and flare states is significant at a level $>3\sigma$.
A similar increase of $N_H$ was noticed 
by \citet{Giardino+2007}  in the flare observed in DROXO and by \citet{Kamata1997} in
a flare observed with ASCA. Such a behavior suggests that the X-rays from the flaring region cross material 
optically thicker than the gas crossed by X-rays coming from the quiescent corona. \citet{Kamata1997}
attributed the increase of $N_H$ to the disk and envelope geometry surrounding \eltn.
They proposed that the flaring sites are preferentially at a low latitudes and their lines of sight crosses
the disk.  This explanation however remains at odds with the face-on geometry of the disk inferred from far IR 
observations \citep[cf.][]{Boogert2002}.

{ We remark that the flare temperature peaks at segment 2, but the EM is detected at its maximum during
segment 3. The time delay between the temperature and the EM peaks is predicted by models of 
flaring loops (e.g., \citealp{Reale2007}): 
the flare heat pulse drives a strong plasma flow from the chromosphere upwards 
along the magnetic tube, and the flow continues to fill the tube for some time after that the heat pulse has stopped 
(and the cooling starts). It is then reasonable to work in the assumption that the flare occurs in a single flaring loop, 
and to use the related diagnostics to determine the characteristics of the flaring loop based on hydrodynamic 
simulations and calibrations on X-ray solar flare observations \citep{Reale2007}.}
In particular, we may infer the semi-length of the loop $L$  from the decay time of the flare, the peak temperature and the slope 
of the decay in the density-temperature diagram, by using equations A.1, A.2 and A.3 in \citet{Reale2007}.
We derive a maximum temperature at the peak of $\sim325$ MK from the $kT$ at segment 2 (11.1 keV $\sim130$ MK) and $\log$ 
EM[cm$^{-3}$]$\sim 54.24$. Because of the large uncertainties in the temperature we cannot derive a reliable value of the slope in
the density-temperature diagram. Thus, we assumed the maximum value of the slope determined by \citet{Reale2007}. 
The maximum slope corresponds to the case of absent sustained heating 
during the flare decay, i.e., consistent with the pure cooling of a single flaring loop. A shallower curve in the density-
temperature diagram would instead suggest the progressive involvement of more and shorter loops, like in arcade flares. Our 
assumption implies that we are deriving an upper limit for the length $L$ of the flaring loop(s).
We measured an $e$-folding decay time $\tau\sim7.65$ ks from the light curve in the soft band 0.3--5 keV (Fig. \ref{flare_timing}) 
and we  estimated $L \leq 2.0\times10^{11}$~cm or $L \leq <2.9 R_\odot$ (or $<0.5$ stellar radii). 

Keeping the same assumption an alternative estimate can be derived from the rise phase, as described in \citet{Reale2007}, 
and in particular from Equation~(12) therein:

\begin{equation}
L \approx 10^9 \psi^2 T^{1/2} \Delta t_R 
\label{eq:trise}
\end{equation}
where $L$ is in $cm$, $T = 332$ MK is the loop maximum temperature at the flare peak, and 
$\Delta t_R$, in $ks$, is the time range between the flare start and its peak (we assume that the peak of the 
light curve is a good proxy of the emission measure peak). From the light curve we measure 
$\Delta t_R = 3 \pm 0.3$~ks. The factor $\psi$ is the ratio between the maximum temperature and the 
temperature at the density maximum. This is unconstrained in our case, and the whole possible range $1.2 < 
\psi < 2$ reported in \cite{Reale2007} is the best we can take for the estimate. From Eq.~(\ref{eq:trise}) we 
obtain a range $0.5 \times 10^{11} < L < 1.3 \times 10^{11}$~cm, i.e, $0.7 R_\odot < L < 1.8 R_\odot$ 
(still $L<0.3$ stellar radii), consistent with the upper limit from the decay time but more stringent.

The energy of the flare is estimated integrating the luminosity during the flare and it is about 
$8\times10^{34}$ erg released in about 20 ks. { Still in the framework of a single flaring structure, and
assuming} a representative semi-length $L \approx 10^{11}$~cm ($\sim 1.5 R_\odot$), a typical cross section 
radius of $R_L = 0.1L$ (see, e.g., \citealp{Golub1980} and \citealp{Klimchuk2000}) and considering the values of EM derived from the 
spectral analysis, we can {push our analysis} to infer a value of the electron density in the loop during the flare from $EM\sim n_e ^2 V$, where $V = 2 \pi R_L^2 L$ is the total volume of the loop.
{Although this is to be taken with care}, for a peak value of EM $\sim2\times10^{54}$ cm$^{-3}$ and $V\sim 6 \times10^{31}$ cm$^3$ we obtain $n_e\sim 2 
\times10^{11}$ cm$^{-3}$, which is similar to the typical values found for solar flaring loops (e.g., 
\citealp{Reale2014}).  %

From the density and temperature, we can in turn infer a minimum strength of the magnetic field ($B_0$) capable to 
confine the plasma inside the loop \citep[e.g.,][]{Maggio+2000}, and this is of order of $B \geq B_0 =  (16 \pi n_e 
k_B T)^{1/2} \sim 500$ ~G, similar to other flares of active stars  and compatible with average fields of 
kG on the stellar surface, as found in other YSOs.

For the second flare, observed only with \nustar, we had {even more} limited information. From the analysis
of the FPM A and B spectra from the flare interval as a whole and from the rise plus peak segments
we derived a plasma temperature of about 5 keV (90\% confidence level: 4.3--6.5 keV) and a $N_H\sim 1.2\times10^{23}$ 
(0.8--1.7 $\times10^{23}$) cm$^{-2}$. The estimate of $N_H$ is less precise than the one inferred from 
\xmm\ spectra as the band below 3.0 keV is not observed by \nustar. However, in agreement with the first flare and its 
previous flares, this second flare of \eltn\ showed once again a $N_H$ value higher than the one measured during the quiescence. 
We will discuss this finding in Sect. \ref{discussion} speculating about the location of the flaring regions in \eltn.  

\subsection{\nustar\ spectroscopy}
\begin{figure*}
\begin{center}
\resizebox{1.01\textwidth}{!}{
 \includegraphics{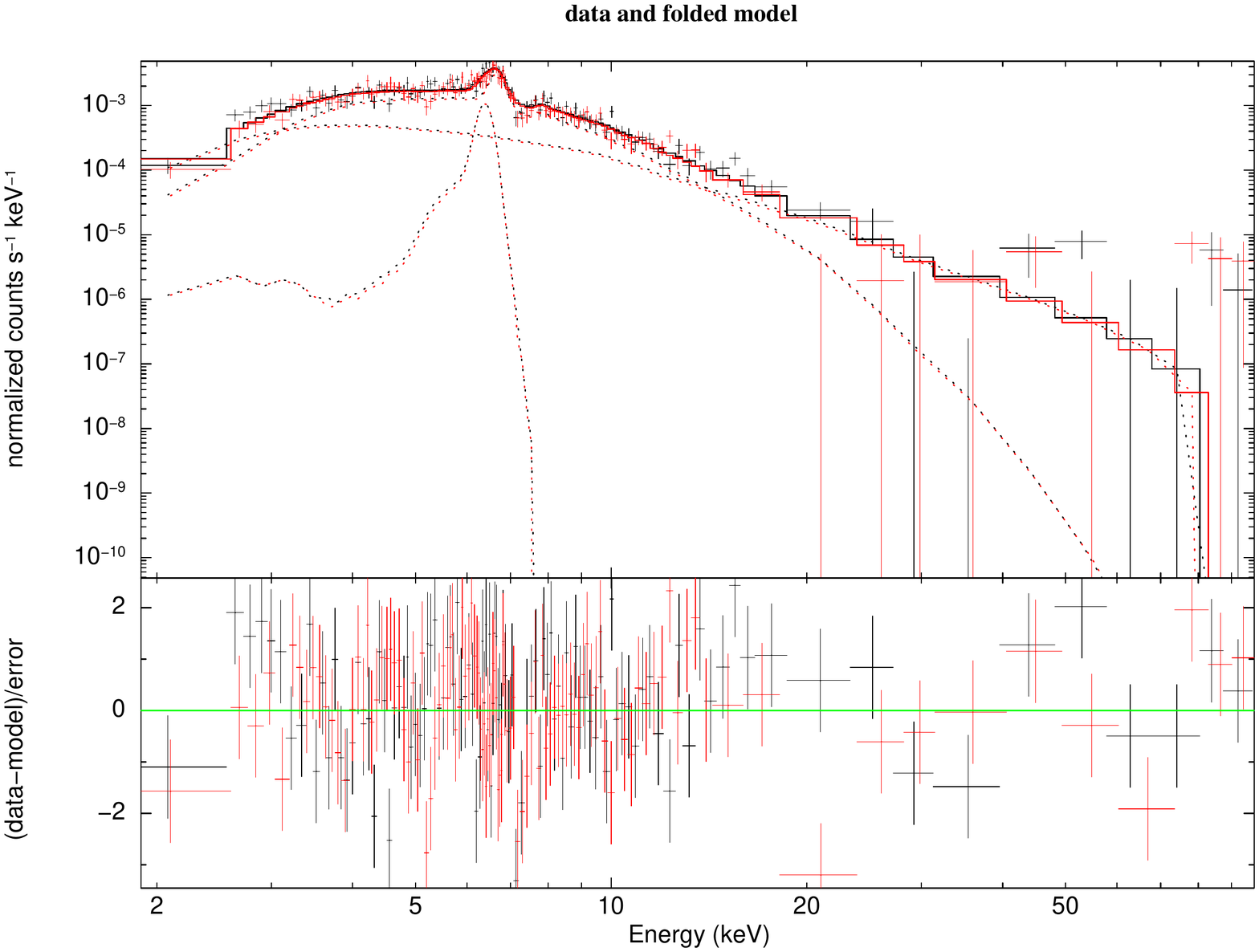} 
 \includegraphics{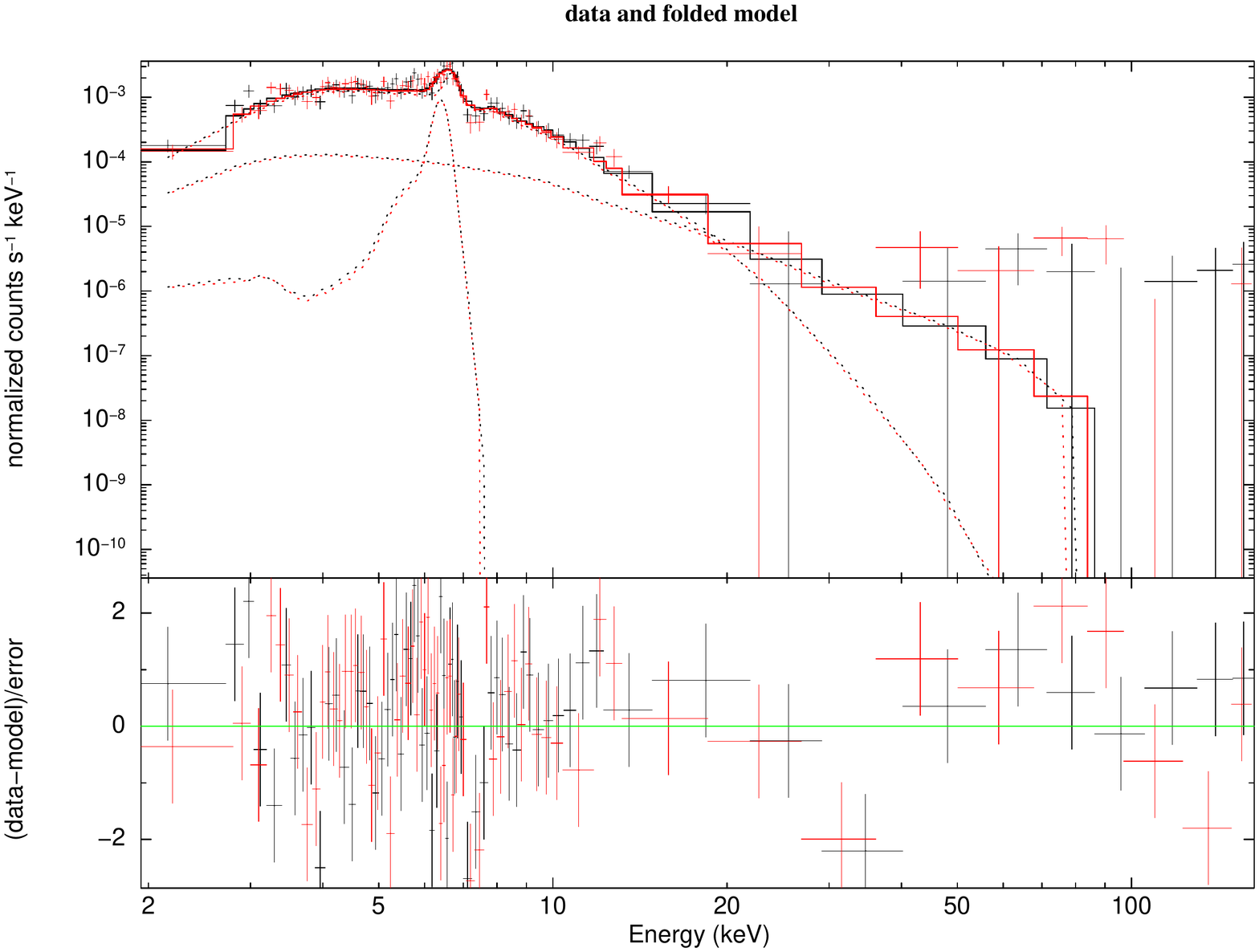} 
 \includegraphics{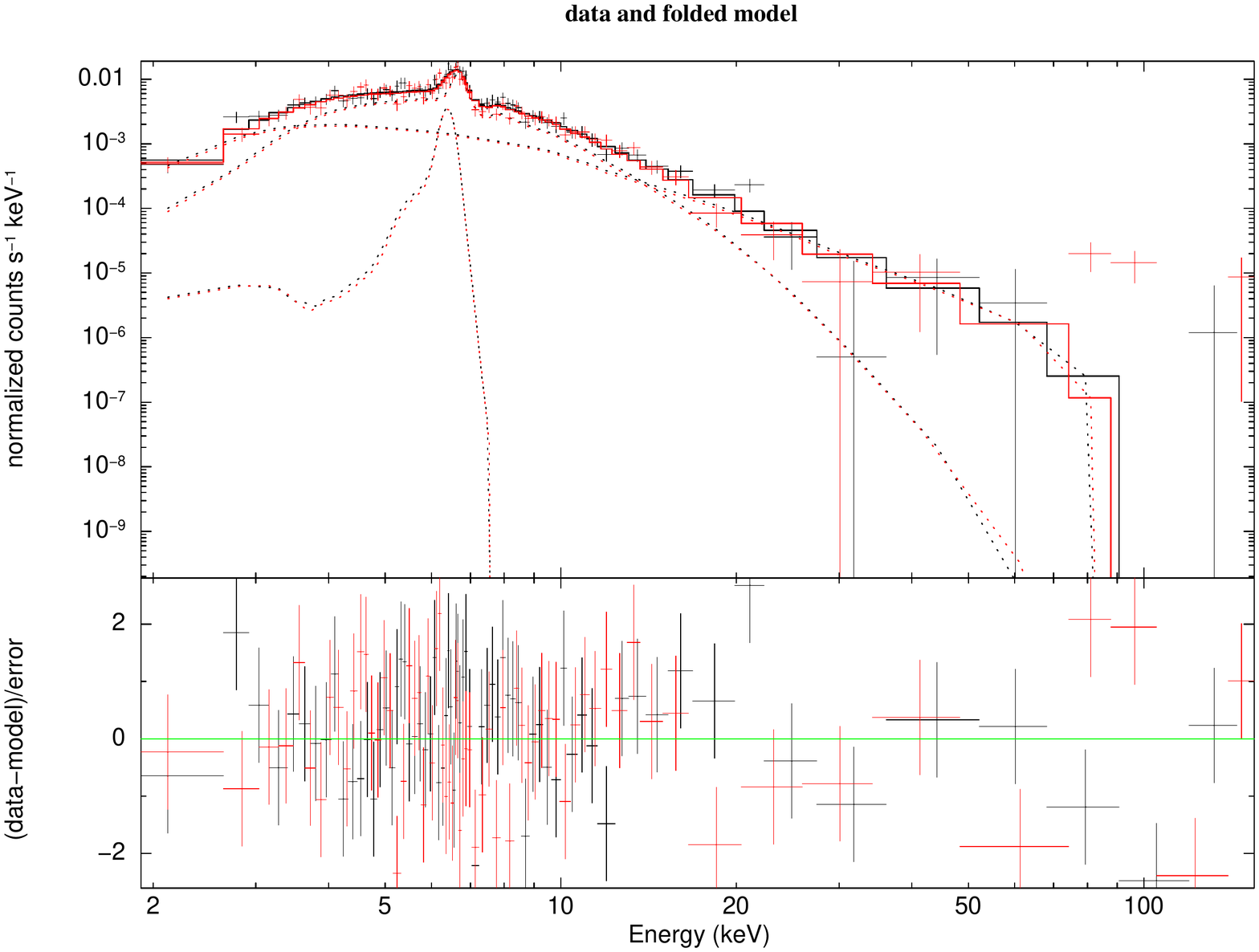} 
}
\end{center}
\caption{\label{nustar_all} \nustar\ FPM A and B spectra in three different time intervals, black is FPMA, red is FPMB.   
The best fit models (dashed lines) and the $\chi^2$ terms in units of $\sigma$ (bottom panels) are also shown.
Left panel: average \nustar\ spectra accumulated on the total exposure ($\approx260$ ks), 
Central  panel: spectra during the quiescent phase. Right panel: spectra during the two flares.}
\end{figure*}

In Fig. \ref{nustar_all} we show the time averaged \nustar\ spectra of the FPMA and FPMB instruments in three different
time intervals. The spectra refer to the total exposure, the sum of the quiescent intervals and the sum 
of the flaring intervals recorded in \nustar\ data. 
We obtained about 4200 net counts in the total spectra,  about 2400 net counts in the quiescent spectra and
about 2000 net counts in the flare spectra.
We plot also a best fit model {where, in this case, we added a power law to the APEC+Gaussian model} for modeling the 
spectra above$\sim10$ keV.
Because of the limited count statistics and energy resolution of \nustar\ FPM instruments, a free centroid 
of the  Gaussian line does not improve the model results and thus we kept fixed the line centroid at 6.4 keV.
A good fit with a model composed by a thermal APEC component and a Gaussian line is satisfactory up to $\sim20$ keV.
{A joint fit of \nustar\ and PN spectra found a temperature similar to that found with the best fit to the 
PN spectrum alone.}
At energies above 20 keV a systematic residual emission is observed in excess of the thermal emission. 
The spectrum has a low statistics in this spectral range, nevertheless the excess is significant above $2\sigma$  
and represents to date the best example of hard X-ray spectrum of a YSOs. 
{Adding a second APEC or a Bremsstrahlung components does not improve the fit above 30 keV as evaluated with 
the $\chi^2$ statistics.
Adding a power law component improves the fit and gives a spectral index 
$\gamma\sim1..5-2$ in the energy band $\sim20-80$ keV. However, the low count statistics above 50 keV 
after background subtraction makes the best fit procedure and the $\chi^2$ test not applicable in the $50-80$ keV range.
We speculate that the emission in $20-50$ keV and modeled with a power law could be of non-thermal nature from 
a population of high energy particles that can contribute to pumping up the fluorescent emission as discussed by 
\citet{Emslie+1986} for energetic solar flares.
}

The excess of hard X-ray emission with respect to the thermal emission is 
detected not only during the flare but also during the quiescent phase. 
This means that the non-thermal component is weak but present also during the quiescent
phase rather than being emitted exclusively during the flare. 
The flux in $10-80$ keV varies in $1.8\times10^{-13}-3.5\times10^{-13}$ \fxu between the quiescent and flaring phases.

\subsection{Fluorescent emission} \label{fluorescence}
Table \ref{fit5-8} reports the values of the centroids, the equivalent widths and the fluxes of the
Gaussian line for the fluorescent emission at $\sim6.4$ keV. We used only the band 5--8 keV as described
in Sect. \ref{observations} to better determine the centroid and the strength of the Gaussian line. 
Fluorescence from cold Fe is present in the spectra of \eltn\ in quiescent and flaring states as detected 
before by \citet{Giardino+2007} and \citet{Favata+2005} and it is variable in strength and in centroid position. 
The spectral region in $\sim6.4-7$ keV is rich of K$\alpha$ and K$\beta$ lines from neutral to multiply ionized Fe 
(see \citealp{Kallman+2004}). {In a few cases the centroid is at 6.5 keV with a 90\% confidence range of $\sim 6.4-6.7$ 
keV and this can be explained by emission  from Fe at higher ionization stages  (cf. \citealp{Emslie+1986}). }

\begin{figure*}
\resizebox{\textwidth}{!}{\includegraphics{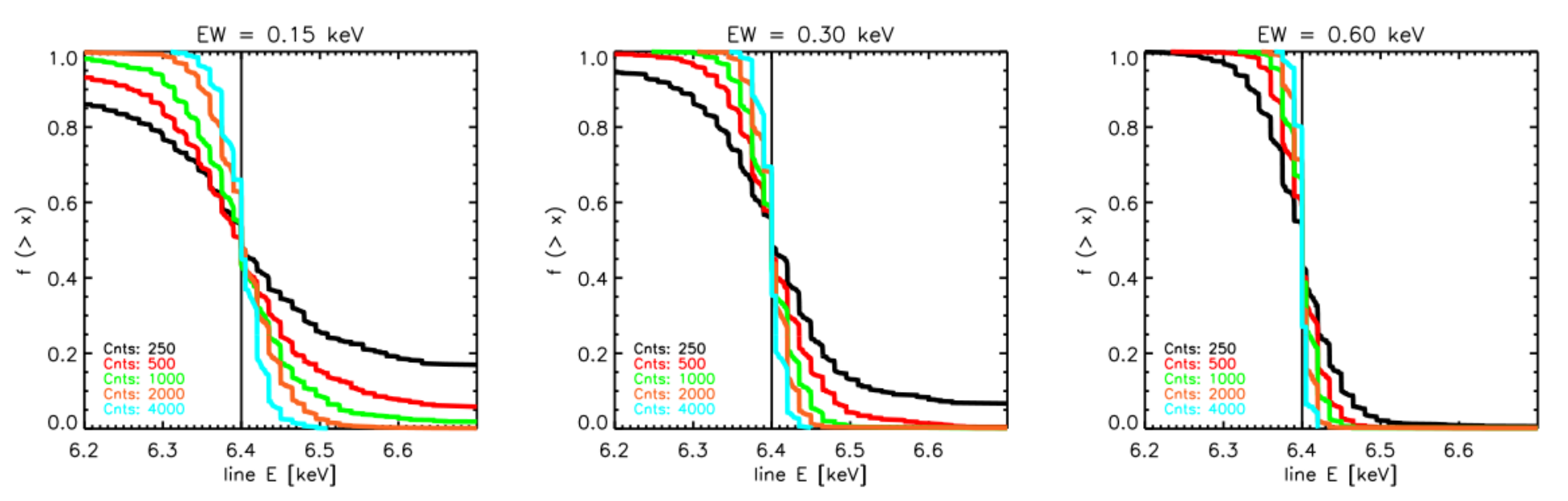}}
\caption{\label{sims} Cumulative distributions of the best fit centroid positions of the Gaussian line 
from the simulations at different levels of count statistics (values in the plots) 
and for three values of the equivalent width of the Gaussian line used in the starting model 
(indicated in the title of the plots).}
\end{figure*}

\begin{figure}
\resizebox{\columnwidth}{!}{\includegraphics{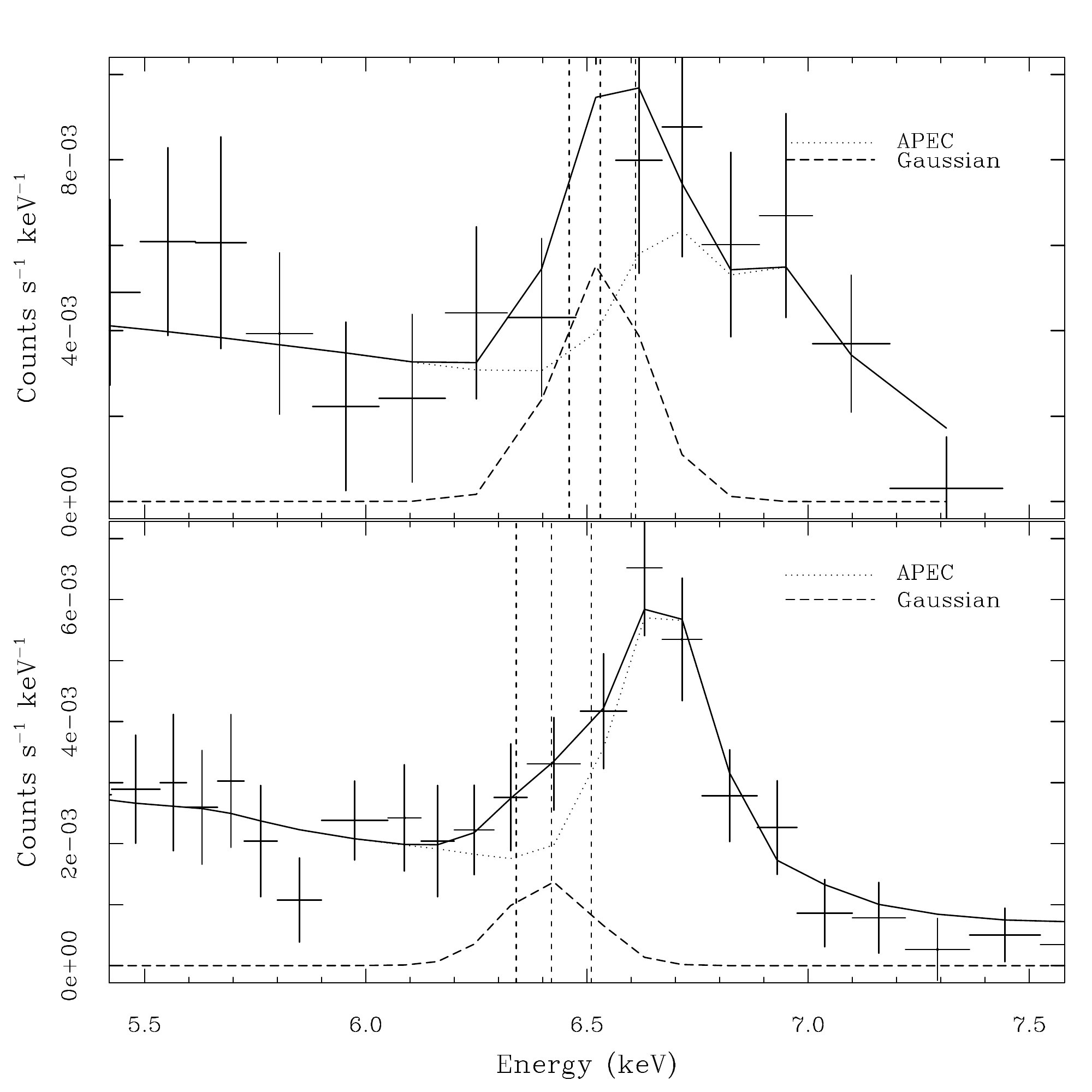}}
\caption{\label{spec6.4kev} Top panel: PN spectrum and best fit model (solid line) during segment 1 of the second
\xmm\ exposure. Dotted lines show the APEC component, dashed lines show the Gaussian component, respectively .
Bottom panel: the same plot for the PN spectrum during the third \xmm\ exposure. Vertical dashed lines  in both panels
mark the centroid positions and the 90\% confidence range.}
\end{figure}

In order to test the robustness of the centroid determination, we made several sets of simulations
of spectra at different levels of count statistics. The spectra are generated from a model composed by a thermal APEC 
component at 4 keV absorbed by a gas column density of $N_H=5.5\times10^{22}$ cm$^{-2}$, with a global abundance $Z\sim0.5 
Z_\odot$ and a Gaussian line at 6.4 keV with equivalent width in the set: 0, 0.15, 0.3, 0.6 and 0.8 keV. 
The abundance was kept fixed in one set of 1000 simulations and variable in another set of 1000 simulations.  
Fig. \ref{sims} shows the cumulative distributions of the line centroid as a function of the count statistics of 
the input spectra and different line intensities.
From these simulations we infer that in the case of spectra with more than 500 counts and with $EW\ge0.3$ keV 
there is a very low probability ($P<0.05$) to determine the centroid at energies well above 6.5 keV. 
In the data, the most significant case where we measured the centroid at 6.51 keV before the flare 
(second \xmm\ exposure, segment 1), 
when the number of counts in the spectra are 476, 76 and 50 for PN, MOS1 and MOS2, respectively, 
and the EW is about 0.47 keV (0.25--0.8 keV 90\% confidence range). 
Here we possibly caught fluorescent emission from partially ionized Fe just at the beginning of the flare. 
Fig. \ref{spec6.4kev} shows the PN spectra during the segment 1 of the second \xmm\ exposure and the third exposure.
  The best fit model is shown and the centroid in one case is found at 
$\sim6.53$ keV and in another case is found at $\sim6.42$ keV.  These values are marginally compatible at the 90\% 
significance level as determined from the uncertainties calculated with XSPEC. Our simulations 
suggest that these two values of the centroid are different at a 95\% significance level given the  
counts of the spectra are $\ge 500$. 

\begin{figure}
\resizebox{\columnwidth}{!}{\includegraphics{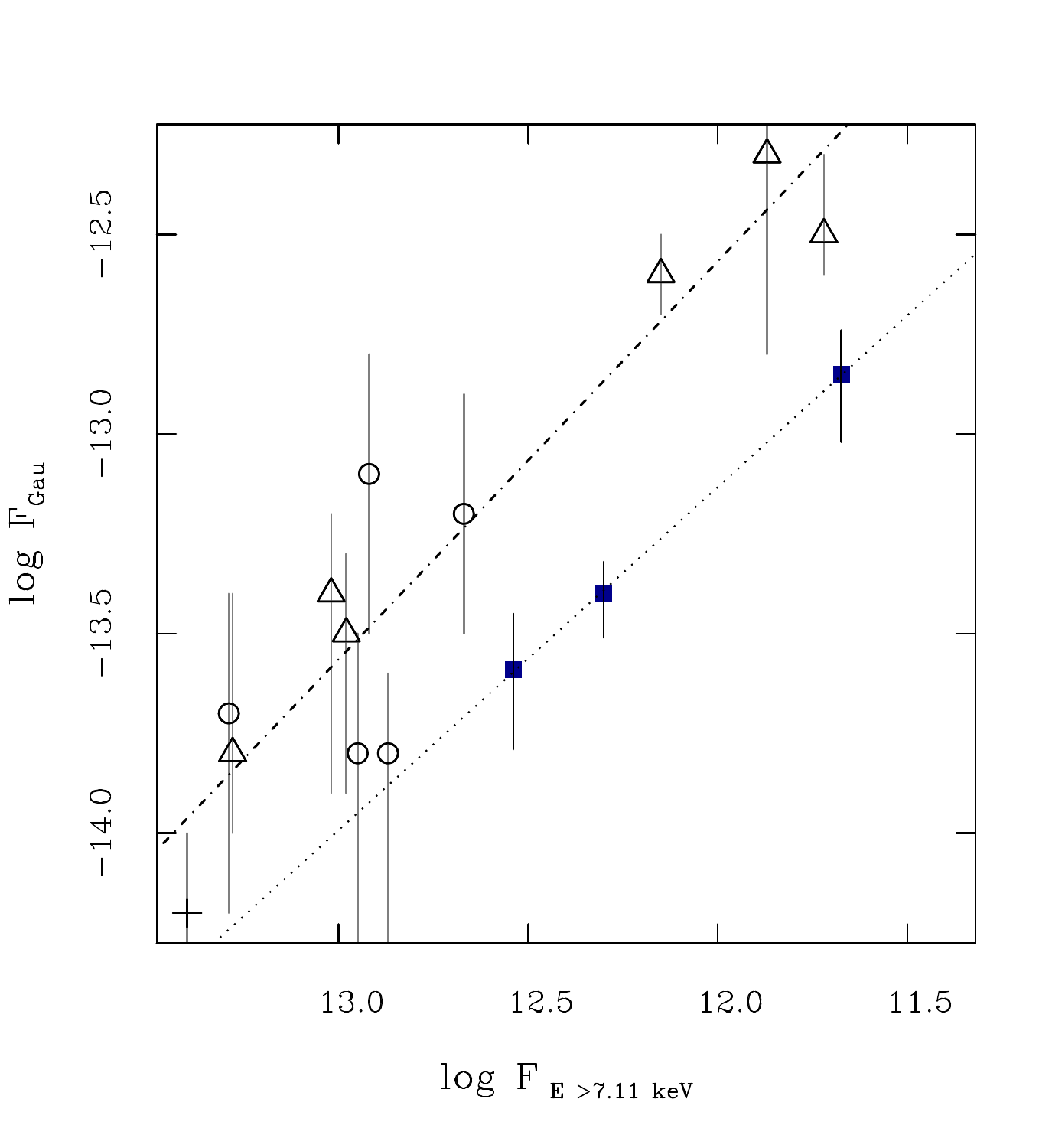}}
\caption{\label{fgauf7.1} Flux of the Gaussian spectral component vs. the flux above 7.11 keV. 
The different symbols refer to the first \xmm\ observation (circles), the second one during the flare (triangles), 
and the  last \xmm\ observation (cross).  The filled symbols refer to the fluxes derived from \nustar\ best fit 
models (see Table \ref{nustarfit}). The lines are the linear best fit to the \xmm\ and \nustar\ data respectively. 
}
\end{figure}

There is marginally significant variability of the line EW during the quiescent intervals just before the first flare, and the other quiescent phases during the first and the third \xmm\ observations in absence of evident flaring activity.  
The EW is found to be between $\sim$0.15 and $\sim$0.47 keV. 
 
On the other hand, the flux of the Gaussian line increased when the overall X-ray flux increased. In particular there is a
correlation between the flux above 7.11 keV and the flux of the line. In Fig. \ref{fgauf7.1} we show a scatter
plot of the flux of the Gaussian line vs. the flux above 7.11 keV. In details,
we calculated the fluxes in the 7.11-10.0 keV band for the \xmm\ spectra with best fit parameters in Table \ref{bestfit} 
and in the 7.11-80.0 keV band for the \nustar\ spectra with best fit parameters in Table \ref{nustarfit}.
A systematic excess of the \nustar\ fluxes in the 7.11-80.0 keV band is present with respect to the 7.11-10.0 keV \xmm\ 
fluxes due to the larger bandwidth of \nustar\ fluxes.
A linear fit between the two fluxes gives a slope of $0.97\pm0.13$ and
intercept of $-1\pm1.6$. Fitting a relationship of the type $y = A x$ gives a slope of $1.044\pm0.005$
The three \nustar\ points gives a slope of $0.86\pm0.02$  and $1.09\pm0.005$ when fitting without intercept. 
The correlation between Gaussian flux and flux above 7.11 keV can be understood by the fact that 
the photons at energies above 7.11 keV can induce Fe fluorescence.
However, it is presumable that very hard X-ray photons {will not be absorbed by material with $N_H < 10^{24}$ }
cm$^{-2}$, thus the photons that concur to excite fluorescence have energies well below 80 keV. 

\section{Discussion}
\label{discussion}
Fluorescent emission is a feature of \eltn, both in the quiescent and flaring states. An EW in excess of 0.15 keV 
is detected in almost all the the time intervals. This result makes unrealistic a simple model made of an irradiated disk, 
and hints that other mechanisms of reverberation and/or a more complex geometry that takes into 
account the cavity where \eltn\ sits can have a role for explaining such high EWs. 
Fe fluorescence in YSOs of ONC has been investigated by \citet{Czesla+Schmitt2010}, they remarked how
explaining the origin of the fluorescent line at 6.4 keV in a few case of quiescent sources is still an open issue.
Their sample of COUP sources span a range of N$_H$ in $2\times10^{20}-2\times10^{23}$ cm$^{-2}$ and EWs between $\sim0.1$ 
and 0.8 keV (including the quoted uncertainties), with the only exception of V 1486 Ori (COUP \# 331) 
which showed EW$\ge1.4$ keV. 

In \nustar\ spectra we detected an excess of hard X-ray emission in \eltn\ likely of non thermal origin. 
The counting statistics do not allow to perform in depth a time resolved 
analysis, yet there is no evidence of an increase or a concentration of such a hard emission  
during flares only as the hard X-ray emission seems produced ubiquitously during the entire observation. 
We speculate that the excess of hard X-ray emission is associated to a population of accelerated particles 
 moving along the accretion streams and varying with stochastic frequency in time due to a highly 
structured magnetic field. 
The average strength of the magnetic field is expected to be of order of a few hundreds of G 
in order to constrain plasma at an average temperature of a few keV, while, in comparison, the average coronal field in the
Sun is of order of 2 G. Locally the magnetic field of YSOs like \eltn\ can reach up to few kG of strength in the core of 
active regions and during flares. Still, it is possible to use the Sun as a template for the corona to build up magnetic 
fields with values in excess of a kG (see \citealp{Orlando2003}).
In such a scenario part of the flux of the Fe K$\alpha$  6.4 keV line could originate from collisions of these 
particles with the disk and a correlation with the intensity of the fluorescence and the non-thermal emission
would be expected. 

\begin{figure}
\resizebox{\columnwidth}{!}{\includegraphics{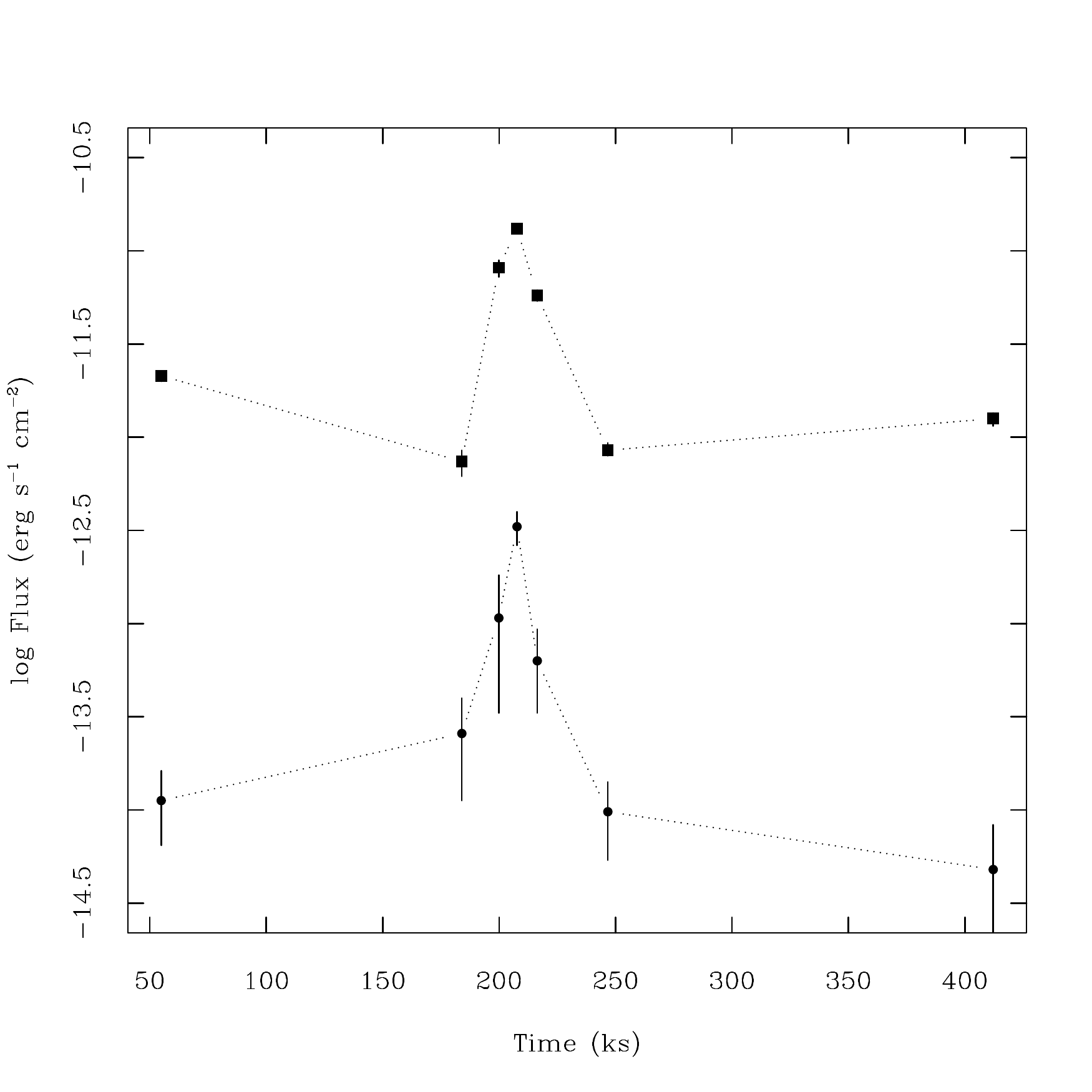}}
\caption{\label{fxfgau} Flux of the APEC thermal component (large symbols) and of the Gaussian line (small dots) 
as a function of the time. Error bars refer to 90\% significance level of uncertainty.}
\end{figure}

However, a clear correlation between the flux of the thermal component and the flux of the Gaussian line
that models the fluorescent emission is still observed during flares. 
In Fig. \ref{fxfgau} we plot the flux of the Gaussian line and the flux of  the APEC component as a function of the time. 
During the flare we observe an increase of both fluxes. We interpret this behavior as the
increase of fluorescence during the flare due to photons with energies above 7.11 keV.
Before and after the flare the value of the flux of the Gaussian line show very little variation at the 
90\% significance level. 
Still, the origin of fluorescence outside the flares remains puzzling when the origin of the phenomenon 
is explained only in terms of high energy photons {as EWs$>0.15$ keV seems difficult to obtain with
a simple irradiated disc model}.
In the Sun, \citet{Parmar1984} found that the fluorescence observed during solar flares is produced 
almost exclusively by photons at $E>7.11$ keV while there is very little evidence of electron induced
fluorescence in the flares they have analyzed. High energy electrons are efficient at stimulating
fluorescence when their energies are $<25$ keV, whereas the efficiency of hard X-ray photons to stimulated 
fluorescence has a cut off at around 20 keV. 
With the data in hand we cannot detect any delay between the increase of the
fluorescence with respect to the overall coronal flux during the flare. In principle one can expect a delay
if the emitting region is the inner disk and the excitation of the fluorescence takes some time to reach its
maximum and to fade out after the flare.

We find suggestion that the centroid of the fluorescent line could vary in time. 
From our simulations we estimated a significance
of such variation at a 95\% confidence level. The change of line centroid can be explained in a scenario 
where the emission arises from excited material at different 
ionization  stages whose relative contributions (and associated energy of excited emission lines) 
to the overall emission in the $6.4-6.6$ keV  energy band 
do vary with time. Future missions like Athena \citep{Nandra+2013, Sciortino+2013} will provide both a 
collecting area larger than \xmm\ and high spectral resolution  ($2.5-3$ eV up to 7 keV, \citealp{Barret+2016}).
In this respect, \eltn\ looks as the most promising candidate among Class I YSOs given its proximity.
However for observations in the \nustar\ band, its weak hard X-ray flux remains still too faint 
to allow for a more detailed time-resolved analysis.

\citet{Isola+2007} found a significant correlation between the soft X-ray emission 
in the GOES band $1.6-12.4$ keV (mostly of thermal nature), and the hard X-ray emission in the RHESSI bands $20-40$ keV 
and $60-80$ keV (mostly of non-thermal nature) during solar flares. This correlation holds up to 
the most energetic events, spanning about four orders of magnitude in peak flux. 
They showed that the same scaling law holds for the handful of available 
hard X-ray observations of intense stellar flares observed with Beppo{\em SAX} in active stellar binaries 
or zero-age main sequence stars. If the X-ray emission in Class~I/II YSOs is just a scaling up of solar phenomena, 
we expect such a correlation to be valid for very young pre-main sequence stars.
In \eltn\ the flux at the peak of the flare from the PN spectrum of segment 3 in the $1.6-12.4$ keV band 
is $F_S = 1.8\times10^{-14}$ W/m$^2$; 
from the \nustar\ spectrum of the whole flare the flux in $20-40$ keV is $1.16\times10^{-13}$ \fxu\ corresponding 
to $F_H = 5.0\times10^{-7}$ photons cm$^{-2}$ s$^{-1}$ keV$^{-1}$. 
For a direct comparison to the \citet{Isola+2007} results we rescaled these two quantities to a distance of 1 AU 
obtaining a flux of $F_S\approx10.97$  W m$^{-2}$  (1.6-12.4 keV band), and $F_H\sim 305\times10^6$ photons cm$^{-2}$ s$^{-1}$ keV$^{-1}$ (20-40 keV band). 
The relationship of Isola et al. predicts $F_S\sim 12.22$ W m$^{-2}$, which is within a $10\%$ of uncertainty   
from our  measurement of $F_S$. It is evinced that in a Class I object like \eltn\ 
the soft and hard fluxes at the flares show the same scaling law empirically found for the Sun and 
active stars. The flare of \eltn\ can be considered a scaled-up version of powerful solar flares and
similar  to those of active stars on the main sequence.

A further test for the analogy between the flare in \eltn\ and the solar flares was based on the thermal 
flux estimated by \citet{Isola+2007}. 
We used the flux in the $20-40$ keV band measured from \nustar\ spectra and the coefficient of scaling $m$ 
given in Table 3 of \citet{Isola+2007} and corresponding to the temperature of 6 keV (the closest to the one observed in \eltn) 
to estimate the thermal flux in the $1.6-12.4$ keV band, and compared it to the one observed in our PN spectrum. 
The flux we obtain is $F_S \sim 1.8\times10^{-14} * m = 6.51\times10^{-7}$ to be compared to 
$5.0\times10^{-7}$ photons cm$^{-2}$ s$^{-1}$ keV$^{-1}$ which is within 30\% uncertainty from the value estimated
from Isola et al. relationship. This again corroborates the analogy between the flare of a very young corona of
a Class I object (\eltn) and the flares of more evolved active stars.

The Coordinated Synoptic Observations (CSI) observed the YSOs in the $\sim$ 3 Myr old NGC 2264 star forming
region simultaneously with {\emph COROT}, \chandra\ and {\emph Spitzer}. Light curves of tens of Class~I, II and III YSOs 
have been obtained during the $\sim$ 3.5 days CSI campaign and several tens of flares were detected. 
\citet{Flaccomio+2018} showed that the energy released during some of such flares, $E_{\rm flare}$, 
can easily reach values up to $\sim$ 10$^{36}$ erg, i.e. up to 1-2 dex 
above the energy released in the strongest (X100 class) flares ever observed on the Sun. 
The flare in \eltn\ released an energy of about $8\times 10^{34}$ erg and this is in the median of the 
energies of the detected flares in NGC 2264. At the same level of flare energy from studies on other SFRs like ONC 
\citep{Caramazza2007} and Cyg OB2 \citep{Albacete+2007}, a frequency of flares of about $1-2$ Ms$^{-1}$ is expected. 
When considering also the DROXO exposure, \eltn\ showed two similar flares detected in a global \xmm\ exposure of 800 ks
which is in good agreement with the flaring rate derived in other SFR regions. 

The energy released during the flare is high even when compared to the energies released by the most 
powerful solar flares, but it is of the same order of magnitude of the energies of flares of similar YSOs. 
On the other hand, the peak temperature is quite high and 
on top of an already hot temperature during quiescence (185 MK vs. $\sim50$ MK).  
A flare with similar duration was also observed in DROXO \citep{Giardino+2007}. 
A compact size of the hosting {loop(s)} is suggested. 

\begin{figure}
\resizebox{\columnwidth}{!}{\includegraphics{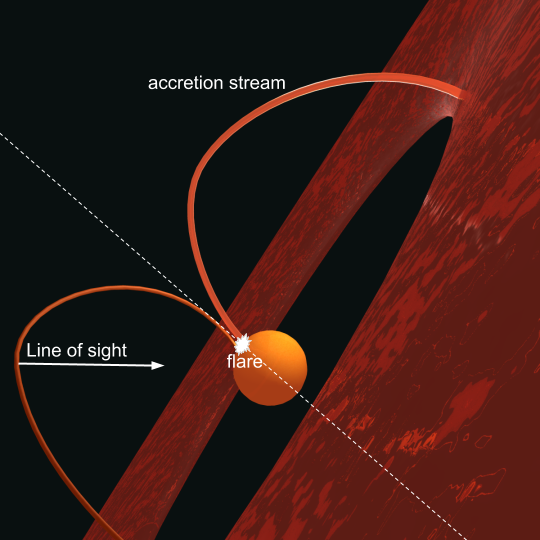}}
\caption{\label{nhcartoon} Cartoon of the proposed scenario to explain the increase of N$_H$ during flares.
The flares originate at the base of the accretion stream located around the pole and, in a face-on geometry,
the emitted X-rays would cross a portion of the accretion stream with N$_H$ absorption larger than
the average $N_H$ value.}
\end{figure}

A peculiar feature of the flares in \eltn\ is the increase of gas absorption of about five times 
during the flares as described in Sect. \ref{flare}. The increase of gas absorption was also observed 
in past flares observed with \xmm\ and ASCA \citep{Giardino+2007,Kamata1997}. 
\citet{Kamata1997} speculated that, under the hypothesis that the disk is edge on, the flares occur at
low latitudes and the X-rays pass through the disk being thus heavily absorbed, 
while the X-rays emitted during quiescent phases are coming from the rest of the corona and encounter less gas 
along the line of sight. However, while plausible, this interpretation conflicts with the
geometry of the source extrapolated from sub millimeter and FIR observations 
(\citealp[e.g.][]{Boogert2002,Ceccarelli2002,Miotello2014,Rocha2015}, see also Fig. \ref{nhcartoon}). 
An alternative explanation that can reconcile the system geometry and the higher gas absorption during flares
is that the flares are generated near the feet of the accretion streams (see Fig. \ref{nhcartoon}. 
The feet could be located preferentially at high stellar latitudes around the stellar poles as the streams follow
the large scale dipolar geometry of the magnetic field and likely these are the sites of frequent 
flares. The X-rays generated during the flares at the stream feet travel a portion of path across the 
dense accreting gas before  arriving to the observer. As a result the gas absorption measured during flares is found
systematically larger than the one measured from the quiescent corona.   

{Finally, a different scenario can in principle explain both the $N_{\rm H}$ enhancement associated to flares and the 
large EW observed for the 6.4 keV line. Elias~29 displays a hard X-ray emission ($E>20 keV$), possibly of 
non-thermal origin, in addition to its thermal X-ray spectrum ($E<20 keV$) of the corona. At these energies the 
number of X-ray photons that undergo to Compton scattering, instead of photo-absorption, could be non negligible. Compton 
scattering diminishes the energy of scattered photons and cause a global softening 
of the X-ray spectrum. Depending on the system geometry, Compton scattering could therefore mimic a $N_{\rm H}$ 
lower than that experienced by the primary X-ray photons emitted by the corona on our line of sight. 
In fact, the line of sight toward Elias~29 passes approximately on the edge of the inner cavity. 
Therefore X-ray photons scattered toward us by the disk surface and the inner cavity wall will experience an $N_{\rm H}$ 
lower than that suffered by primary photons emitted by the central star. Hence, the total X-ray spectrum reaching us 
would be approximately composed of highly-absorbed primary X-ray photons, and less-absorbed scattered photons. In this 
scenario, an increase of the thermal emission of the corona (i.e. a flare) implies an increase of the highly-absorbed 
primary-photon component only. That would therefore explain why the X-ray flares observed on Elias~29 
show $N_{\rm H}$ systematically higher than that of the quiescent phases. 
In addition, assuming that the real $N_{\rm H}$ between us and the central star is that observed during flares (i.e. 
$\sim2\times10^{23}$\,cm$^{-2}$), the EW of the fluorescent line at 6.4\,keV will be larger than the 
model predictions simply  because, at these $N_{\rm  H}$ values, photons at $\sim7.1$\,keV start to be 
significantly absorbed, while the fluorescent line, originating preferentially from the disk 
surface and the inner cavity wall, will be less absorbed. Such a scenario is qualitatively similar to that of AGNs, 
where the EW of the 6.4\,keV line is observed to increase up to values of $\sim1$\,keV for increasing $N_{\rm H}$ 
(e.g. \citealp{Fukazawa2011}). The case of Elias~29, i.e. $EW\sim0.3$ keV and $N_{\rm H}\sim10^{23}$\,cm$^{-2}$, 
would neatly fit the EW vs $N_{\rm H}$ pattern observed for the AGNs.}

\section{Summary}
\label{conclusions}
We presented the results of a joint \xmm\ and \nustar\ simultaneous observation of \eltn, the IR brightest Class I 
object in the Rho Ophiuchi Dark Cloud core F (LDN 1688). 
The full EPIC image contains about 100 X-ray sources while \nustar\ shows about ten sources among which \eltn\ is the 
brightest. We observed a flare of duration of about 20 ks with a regular exponential-like decay characterized by an 
e-folding time of about 7.6 ks in the $0.3-5.0$ keV band and a rise time of about 3 ks. 
Through time resolved spectroscopy we inferred the properties of the quiescent and the flaring plasma. 
We determined that {the flaring structures are relatively compact with a length of about $1-2$ R$_\odot$,} 
which   suggest that likely the loop is anchored to the stellar surface. 
A magnetic field of at least 500 G is required to confine the plasma within the loop.
A second flare with a duration of 50 ks was observed with \nustar\ only and for which we inferred a loop size similar to the
one that hosted the first flare.  
During these flares we observed an increase of $N_H$ that suggests a specific location of the flaring sites
at the base of the accretion streams of the star. 

Fluorescent emission from neutral or partially ionized Fe is observed both during quiescence and during the flares.
Fluorescence is modeled with a Gaussian line varying both in the centroid position and in strength. 
There is a clear correlation between fluorescent emission and coronal emission during the flare. 
 However, there is still significant Fe fluorescence outside the flare that cannot be explained exclusively 
with the contribution of photons at $E>7.11$ keV.  
We detect a hard X-ray emission from \eltn\ above $\sim20$ keV which is not explained by a thermal emission. 
We argue that a non-thermal population of electrons accelerated from the coronal magnetic field could be responsible 
for  this emission.  We speculate that the same population could contribute to the fluorescent emission of 
\eltn. 

\begin{acknowledgements}
We thank the anonymous referee for her/his comments and suggestions which improved the manuscript.
The authors acknowledge modest financial contribution from the agreement ASI-INAF n.2017-14.H.O.
IP acknowledges support from the ASI and the Ariel Consortium.
We made use of $R$, a language and environment for statistical computing, the XMM-SAS suite,
HEASOFT fv, XSPEC and DS9.
\end{acknowledgements}

\begin{landscape}

\begin{table}[ht]
\caption{\label{bestfit} 
Parameters from model best fit in the spectra from  the time intervals identified with the PELT algorithm (Fig. \ref{cpt_pn}). }
\resizebox{1.1\textheight}{!}{
\begin{tabular}{lllllllllllllll}
  \hline \hline 
Obs & Segment & Counts & $N_H$ & kT & log EM & log Flux &  log L$_X$ & Centroid & log F$_{Gau}$ & EW & log F$_{E>7.11 keV}$ & $\chi^2 _{red}$ & Prob & D.o.F. \\
      &      &   & $10^{22}$ cm$^{-2}$ & keV & cm$^{-3}$ & \fxu & \lxu & keV & \fxu & keV & \fxu &  & & \\
  \hline
First & \multicolumn{11}{c}{}  \\ 
   & 1 & 2049+200+190 & 7.2 (4.8--11) & 4 (2.3--8.6) & 53.24 (53.18--53.28) & -11.81 (-12-- -11.48) & 30.43 & 6.63 (6.49--6.75) & -13.1 (-13.72-- -12.94) & 0.21 & -12.92 & 1.07 & 0.3 & 91 \\ 
  & 2 & 2317+662+851 & 7.2 (6.6--7.9) & 4.2 (3.5--5) & 53.32 (53.28--53.33) & -11.73 (-11.79-- -11.67) & 30.51 & 6.48 (6.42--6.55) & -13.73 (-14.03-- -13.56) & 0.23 & -12.87 & 1.14 & 0.14 & 131 \\ 
  & 3 & 1225+369+500 & 6.4 (5.8--7.12) & 5.4 (4.2--6.9) & 53.35 (53.32--53.36) & -11.67 (-11.7-- -11.61) & 30.57 & 6.57 (6.51--6.63) & -13.31 (-13.57-- -13.14) & 0.23 & -12.67 & 0.99 & 0.48 & 75 \\ 
  & 4 & 1074+314+464 & 6.3 (5.6--7.1) & 3.4 (3--4.4) & 53.35 (53.32--53.38) & -11.72 (-11.79-- -11.65) & 30.52 & 6.31 (5.97--6.5) & -13.87 (-14.82-- -13.6) & 0.19 & -12.95 & 1.54 & 0.003 & 67 \\ 
  & 5 & 962+250+401 & 5.4 (4.4--6.2) & 5.4 (3.8--14.1) & 53.03 (53--53.05) & -11.99 (-12.11-- -11.9) & 30.25 & 6.57 (6.5--6.63) & -13.56 (-13.85-- -13.28) & 0.28 & -13.29 & 1.9 & 4e-05 & 59 \\    \hline
 Second & \multicolumn{11}{c}{}  \\ 
  & 1 & 1800+389+351 & 9 (6.8--12.4) & 3.9 (2.6--6.3) & 53.12 (53.08--53.15) & -11.94 (-12.09-- -11.7) & 30.30 & 6.49 (6.4--6.57) & -13.62 (-13.96-- -13.42) & 0.51 & -12.98 & 1.15 & 0.15 & 91 \\ 
  & 2 & 534+141+193 & 21.3 (15.9--27.2) & 11.1 (5.2--64) & 54 (53.95--54.04) & -11.06 (-11.16-- -10.78) & 31.18 & 6.6 (6.38--6.66) & -12.44 (-13.23-- -12.21) & 0.3 & -11.87 & 1.36 & 0.1 & 27 \\ 
  & 3 & 1842+583+660 & 19.9 (18.4--21.5) & 6.9 (5.4--6.8) & 54.24 (54.22--54.28) & -10.69 (-10.74-- -10.65) & 31.55 & 6.4 (6.38--6.43) & -12.39 (-12.49-- -12.3) & 0.36 & -11.72 & 1.65 & 7e-05 & 91 \\ 
  & 4 & 952+355+371 & 15.6 (14.2--17.3) & 5.5 (4.5--7.9) & 53.79 (53.77--53.82) & -11.22 (-11.36-- -11.15) & 31.02 & 6.5 (6.59--6.45) & -12.85 (-13.01-- -12.72) & 0.59 & -12.15 & 1.39 & 0.028 & 55 \\ 
  & 1and5 & 2919+707+714 & 7 (6.1--7.9) & 5.9 (4.9--8.7) & 52.82 (52.8--52.84) & -12.19 (-12.37-- -12.14) & 30.05 & 6.51 (6.46--6.58) & -13.73 (-13.88-- -13.6) & 0.51 & -13.02 & 1.22 & 0.04 & 137 \\ 
  & 5 & 1455+472+509 & 7.2 (6.4--8.1) & 3.6 (3.1--4.3) & 52.94 (52.92--52.96) & -12.16 (-12.24-- -12.09) & 30.08 & 6.56 (6.49--6.61) & -13.71 (-13.89-- -13.57) & 0.56 & -13.28 & 1.36 & 0.015 & 86 \\    \hline
 Third & \multicolumn{11}{c}{}  \\ 
  & 1 & 2040+685+782 & 5.5 (5.02--6.37) & 4.2 (3.5--5.6) & 52.78 (52.77--52.79) & -12.22 (-12.3-- -12.17) & 30.02 & 6.42 (6.34--6.51) & -14.31 (-14.84-- -14.08) & 0.19 & -13.4 & 1.2 & 0.062 & 127 \\ \hline 
 & Quiesc & 3391 &  5.8 (5.3--6.4) & 4.2 (3.6--5.3) & 52.82 (52.75--52.89) & -12.21 (-12.23-- -12.20) &  30.03 & 6.49 (6.40--6.60) & -14.17 (-14.41-- -14.02) & 0.25 (0.16--0.38) & -13.15 & 1.1 & 0.21 & 127\\
  \hline 
\end{tabular}
\footnotesize 
}

Note: the model is an absorbed APEC thermal component plus a Gaussian that models the fluorescence at $\sim6.4-6.6$ keV.
Ranges at the 90\% confidence level are indicated in braces, in some cases the high limit 
of temperature uncertainty is not constrained. A value of $Z/Z_\odot$ of 0.54 has been derived from the third exposure
(quiescent phase) and used for the  best fit of the spectra of the first and second exposures. 
The 90 confidence level uncertainty of the equivalent widths of the Gaussian line is of order of 0.1-0.3 keV.
Unabsorbed fluxes for the APEC and the Gaussian components are given in the energy band 0.3--8.0 keV, 
the fluxes at $E>7.11$ keV are also listed. 

{ X-ray luminosities are calculated from fluxes using a  distance of 120 pc}.

\end{table}
 
\begin{table}[ht]
\caption{\label{nustarfit} Parameters of the best fit models to the \nustar\ spectra shown in Fig. \ref{nustar_all}.}
\resizebox{1.\textwidth}{!}{
\begin{tabular}{lccccccccc}
  \hline \hline
Spectrum & N$_H$ & kT & $Z/Z_\odot$ & log EM & $\gamma$ & EW & log F$_{20-40\ keV}$ & log F$_{7.11-80\ keV}$ \\ 
         & $10^{22}$ cm$^{-2}$ & keV &        & cm$^{-3}$  &    & keV  & \fxu & \fxu  \\
  \hline
 Total & 14.1 (8.9 -- 20.7) & 3.9 (3.2 -- 4.7) & 0.53 (0.42 -- 0.66) & 53.4 (53.27 -- 53.57) & 2.1 (1.7 -- 2.3) & 0.34 (0.25 -- 0.49) & -13.04 & -12.29 \\ 
Quiescent & 8.8 (5.3 -- 14.8) & 3.6 (2.8 -- 4.4) & 0.53 (NA -- NA) & 53.3 (53.19 -- 53.47) & 2.2 (0.6 -- 2.9) & 0.39 (0.27 -- 0.61) & -13.4 & -12.55 \\ 
Flare & 19.4 (13.5 -- 26) & 4.1 (3.5 -- 4.9) & 0.53 (NA -- NA) & 53.96 (53.85 -- 54.1) & 1.9 (1.7 -- 2.2) & 0.28 (0.18 -- 0.45) & -12.36 & -11.64 \\ 
\hline
\end{tabular}
 }
\end{table}
\begin{table}[ht]
\caption{\label{fit5-8} Parameters from model best fit of the spectra in $5.0-8.0$ keV band.   } 
\resizebox{1.1\textwidth}{!}{
\begin{tabular}{lllllllllllll}
  \hline \hline 
Obs & Segment & Counts & kT & EM & Z & Centroid & N$_{Gau}$ & EW & F$_{APEC}$ & F$_{Gau}$ & $\chi^2 _{red}$ & Prob \\ 
    &        &(PN, MOS1, MOS2)& keV & cm$^{-3}$ & & keV & 10$^{-7}$ photons cm$^{-2}$ s$^{-1}$ & keV & \fxu & \fxu & & \\
\hline
  First  & all & 1716+353+537 & 3.6 (2.9 -- 4.2) & 53.39 (53.28 -- 53.52) & 0.53 (0.45 -- 0.61) & 6.42 (6.38 -- 6.48) & 11.4 (6.7 -- 16) & 0.15 (0.10-0.2) & -11.67 (-11.7 -- -11.66) & -13.95 (-14.19 -- -13.79) & 1.22 & 0.11 \\ \hline
  Second & 1 & 463+76+50 & 10.8 (3.5 -- 13.4) & 52.86 (52.67 -- 53.22) & 1 (0.2 -- 5) & 6.53 (6.43 -- 6.61) & 25 (8.4 -- 44) & 0.47 (0.25-0.8) & -12.13 (-12.21 -- -12.07) & -13.59 (-13.95 -- -13.4) & 0.71 & 0.83 \\ 
         & 2 & 330+50+78 & 7.9 (5.3 -- 9.1) & 53.8 (53.68 -- 53.94) & 1.5 (0.9 -- 2.4) & 6.51 (6.4 -- 6.7) & 160 (62 -- 540) & 0.21 (0.10-0.4) & -11.09 (-11.14 -- -11.05) & -12.97 (-13.48 -- -12.74) & 1.05 & 0.39 \\ 
         & 3 & 1239+230+303 & 7 (6.3 -- 7.9) & 54.02 (53.96 -- 54.08) & 1.65 (1.3 -- 2.1) & 6.4 (6.37 -- 6.43) & 320 (260 -- 390) & 0.41 (0.33-0.5) & -10.88 (-10.9 -- -10.86) & -12.48 (-12.58 -- -12.4) & 1.15 & 0.21 \\ 
         & 4 & 515+128+126 & 4.8 (3.2 -- 6.7) & 53.78 (53.65 -- 53.98) & 0.54 (0.35 -- 0.79) & 6.48 (6.43 -- 6.58) & 93 (55 -- 135) & 0.22 (0.15-0.4) & -11.24 (-11.27 -- -11.21) & -13.2 (-13.48 -- -13.03) & 0.69 & 0.87 \\ 
         & 5a & 459+75+100 & 2.1 (1.4 -- 4.1) & 53.35 (52.95 -- 53.74) & 0.75 (0.1 -- 1.3) & 6.33 (6.06 -- 6.6) & 6.3 (1.4 -- 26.7) & 0.21 (0.11-0.5) & -11.74 (-11.78 -- -11.7) & -14.24 (-14.82 -- -14) & 0.47 & 0.98 \\ 
         & 1\&5 & 953+176+152 & 4.3 (1.9 -- 5.9) & 52.95 (52.92 -- 52.99) & 0.6 (0.4 -- 0.9) & 6.45 (6.4 -- 6.52) & 12 (6.5 -- 16) & 0.34 (0.23-0.6) & -12.07 (-12.1 -- -12.03) & -14.01 (-14.27 -- -13.85) & 1.13 & 0.27 \\ \hline 
  Third  & all & 564+126+125 & 2.3 (1.5 -- 3.8) & 53.28 (52.89 -- 53.71) & 0.4 (0.3 -- 0.5) & 6.42 (6.34 -- 6.53) & 4.9 (1.4 -- 8.5) & 0.21 (0.10-0.4) & -11.9 (-11.94 -- -11.87) & -14.32 (-14.9 -- -14.08) & 1.31 & 0.12 \\ 
\hline
\end{tabular}
}

Note: For the first and third \xmm\ observations we  considered the spectra of the full exposure, for the second one we 
considered the spectra in the time intervals identified with PELT (see Fig. \ref{cpt_pn}). Uncertainty ranges at 90\% 
confidence level are in reported. Gas absorption has been fixed to $N_H=5.5 \times10^{22}$ cm$^{-2}$. 
Fluxes of the APEC and the Gaussian components are calculated in the 0.3-8.0 keV band, these fluxes are plotted
in Fig. \ref{fxfgau} as a function of the time.

\end{table}
\end{landscape}


\end{document}